\documentclass[a4paper,fleqn,usenatbib,useAMS]{mnras}
\usepackage{graphicx}	
\usepackage{amsmath}	
\usepackage{amssymb}	
\usepackage{multicol}        
\usepackage{bm}		
\usepackage{pdflscape}	
\usepackage[T1]{fontenc}
\usepackage{ae,aecompl}

\newcommand{\lr}[1]{\left( #1 \right)}
\newcommand{\lrs}[1]{\left[ #1 \right]}
\newcommand{\lrt}[1]{\left< #1 \right>}
\newcommand{\bbe}{{\boldsymbol \beta}}
\newcommand{\bth}{{\boldsymbol \theta}}

\newcommand{\mbA}{{\mbox{\boldmath$A$}}}
\newcommand{\mbe}{{\mbox{\boldmath$e$}}}

\newcommand{\rSN}{{\rm SNR}}

\newcommand{\tmbe}{{\tilde\mbe}}
\newcommand{\Dmbe}{{\Delta \mbe}}
\newcommand{\Dtmbe}{{\Delta \tilde\mbe}}

\newcommand{\mbetrue}{{\mbe_{g}}}

\newcommand{\ReS}{{S^{Re}}}
\newcommand{\ReN}{{N^{Re}}}

\newcommand{\tilG}{{\tilde G}}
\newcommand{\tilI}{{\tilde I}}
\newcommand{\tilP}{{\tilde P}}

\newcommand{\tilS}{{\tilde S}}
\newcommand{\tilN}{{\tilde N}}
\newcommand{\tilR}{{\tilde R}}
\newcommand{\tilReS}{{\tilde S^{Re}}}

\newcommand{\tilW}{{\tilde W}}

\newcommand{\cM}{{\cal M}}
\newcommand{\cS}{{\cal S}}

\newcommand{\MOM}[2]{{\cM^{#1}_{#2}}}

\newcommand{\SHP}[2]{{\cS^{#1}_{#2}}}

\newcommand{\CoefW}{{C_W}}
\newcommand{\CoefP}{{C_P}}
\newcommand{\eqG}{{\doteq}}
\newcommand{\RWRW}[1]{{Q\lr{#1}}}

\usepackage{newtxtext,newtxmath}

\title[Analytical Noise Bias Correction for Weak Lensing Shear Analysis with ERA]{Analytical Noise Bias Correction for Weak Lensing Shear Analysis with ERA}

\author[Yuki Okura, Toshifumi Futamase]
{Yuki Okura,$^{1,2}$\thanks{E-mail: yuki.okura@riken.jp}
Toshifumi Futamase,$^{3}$
\\
$^{1}$PIKEN\\
$^{2}$NAOJ\\
$^{3}$Kyoto Sangyo University}
\begin{document}
\label{firstpage}
\pagerange{\pageref{firstpage}--\pageref{lastpage}}
\maketitle
\begin{abstract}
Highly precise weak lensing shear measurement is required for statistical weak gravitational lensing analysis such as cosmic shear measurement to achieve severe constraint on the cosmological parameters. For this purpose, the accurate shape measurement of background galaxies is absolutely important in which any systematic error in the measurement should be carefully corrected. One of the main systematic error comes from photon noise which is Poisson noise of flux from the atmosphere(noise bias). We investigate how the photon noise makes a systematic error in shear measurement within the framework of ERA method we developed in earlier papers and gives a practical  correction method. The method is tested by simulations with real galaxy images with various conditions and it is confirmed that it can correct $80 \sim 90\%$ of the noise bias except for galaxies with very low signal to noise ratio. 
\end{abstract}

\onecolumn
\section{Introduction}
It is widely recognized that weak gravitational lensing shear analysis is a unique and powerful tool to analyze the mass distribution of the universe.
Coherent deformation of the shapes of background galaxies carries not only the information of intervening mass distribution but also the cosmological background geometry and thus the cosmological parameters (Mellier 1999; Schneider 2006; Munshi et al. 2008).

In particular, the cosmic shear which is the weak lensing by the large scale structure of the universe has attracted much attention for the measurement of dark energy which is believed to be the reason of accelerated expansion of the universe. 
Although the cosmic shear have been measured by several groups in the past(Bacon et al. 2000, 2003; Maoli et al. 2001; Refregier et al. 2002; Hamana et al. 2003; Casertano et al. 2003; van Waerbeke et al. 2005; Massey et al. 2005; Hoekstra et al. 2006), the accuracy of the measurements are not enough to give a useful constraint on the dark energy parameter(which includes the dark energy equation of state parameter $w$). 
For more precise measurements, several wide survey observations were started such as 
Hyper Suprime-Cam(HSC : http://www.naoj.org/Projects/HSC/HSCProject.html),  
Kilo-Degree Survey(KiDS : http://kids.strw.leidenuniv.nl/),  
The Deep Lens Survey(DLS : http://matilda.physics.ucdavis.edu/working/website/index.html), 
Canada-France-Hawaii Telescope Legacy Survey(CFHTLS : http://www.cfht.hawaii.edu/Science/CFHTLS/)
Dark Energy Survey(DES : https://www.darkenergysurvey.org/), 
and are planned such as
The Large Synoptic Survey Telescope(LSST : http://www.lsst.org), 
EUCLID(EUCLID : http://sci.esa.int/euclid),  
Wide Field Infrared Survey Telescope(WFIRST : https://wfirst.gsfc.nasa.gov/). 

For example, the Hyper Suprime-Cam Subaru Strategic Program(SSP) plans 1400 deg$^2$ wide survey observation for constraining the cosmological parameters with lower than 1$\%$ error.
The observation has started in 2014, and  100deg$^2$ of HSC wide survey data was recently published.
To achieve a severe constraint on the dark energy equation of state parameter for dark energy, 
the HSC SSP requires highly precise weak gravitational lensing shear analysis method with lower than $1\%$ systematic error.

Various methods of weak lensing shear analysis have been developed for this purpose 
such as Bayesian Fourier domain method(Bernstein 2014),
Auto convolved method(Li and Zhang 2016), and 
Metacalibration method(Sheldon 2017, Huff 2017) 
Some of them were tested with realistic simulation(Heymans et al 2006, Massey et al 2007, Bridle et al 2010 and Kitching et al 2012).
There are some systematic biases in the weak shear measurement and they should be accurately 
corrected for any method. 
Among them, the photon noise bias which is bias due to random count from Poisson noise of sky brightness is known to be critical for the  measurement of cosmic shear.  
Comments, investigation and development about the bias can be seen in Bernstein \& Jarvis 2002, Hirata et al. 2003, Miller et al. 2007, Bridle et al. 2010, Refregier et al. 2012, Melchior \& Viola 2012, Kacprzak et al. 2012, Bernstein \& Armstrong 2014, Jarvis et al 2016, Bernstein et al 2016, Simon \& Schneider 2016, Herbonnet et al 2017, Conti et al 2017, Hall \& Taylor 2017, Hoekstra 2017. 

We have also developed a new weak lensing shear analysis method called EHOLICs 
 (Okura and Futamase 2011, 2012, 2013) based on 
KSB(Kaiser et al. 1995) method, 
The method uses an elliptical weight function in order to avoid for the systematic error caused by the approximation of the weight function. 
We have then developed a new method of PSF correction called   
ERA (Okura and Fuatase 2014, 2015, 2016). 
The method re-smears galaxy image and Point Spread Function(PSF) image to have an idealized PSF 
which has the same ellipticity with true ellipticity, Although we have shown that ERA was able to eliminate some systematic error with enough precision, the systematic error caused by the photon noise(noise bias) is not yet corrected. 
Since the photon noise is the Poisson noise of sky count, it is the random count on the observed galaxy image. 
Although it is random, the change of ellipticity of the galaxy image  has not only a random component 
but also has a systematic component which brings about the systematic error in measuring ellipticity. 

In this paper, we investigate the photon noise from the first principle in the framework of ERA method 
and develop its correction in order to achieve required accuracy for cosmic shear measurement.  

This paper is organized as follows.
In section 2, we give a brief introduction of the basics of the ERA method.
In section 3, we derive how to correct photon noise effect analytically based on the ERA method. 
In section 4, we test the correction method by simple simulation.
In section 5, we summarize and discuss our results.

\section{The Basics of The ERA Method}
The ERA method provides a new method of PSF correction by introducing 
Re-Smearing function(RSF). The RSF re-smears the observed galaxy and star images again to reshape PSF into the idealized PSF which has the same ellipticity with the lensed galaxy before the atmospheric smearing.   
The detailed explanation of ERA method can be seen in Okura and Futamase 2016.

\subsection{The notation and definitions}
In this section, we introduce some basic of the lens mapping, the definitions of image moments, SNR, and the 
radius of galaxy images.

The lensing distortion can be regarded as a mapping between the image plane and the source plane. 
On the other hand, ERA method focuses on the change of the ellipticity directly and introduces a plane called the zero plane for individual background galaxies in which galaxy has no ellipticity. 
Thus we regard the intrinsic ellipticity $\mbe_z$  in the source plane as the result of a mapping from the zero plane. 
The observed ellipticity  $\mbetrue$  are the result of this mapping and the lensing shear $\mbe_s$ as follows
\begin{eqnarray}
\label{eq:nonle}
\mbetrue &=& \frac{\mbe_z + \mbe_s}{1 + \mbe_z\mbe_s^*}.
\end{eqnarray}
Since we concentrate on the accurate measurement of the observed ellipticity ``$\mbetrue$'' without any systematic bias, we consider the image plane and the zero plane.   We denote the position 
angle in the image plane by ``$\bth$'' with the origin at the image centroid
and denote the position angle as ``$\bbe$'' in the zero plane with the origin at the image centroid.
The positions in two planes are related each other as   
\begin{eqnarray}
\bbe &=& \lr{1 - \kappa}\lr{\bth - \mbetrue\bth^*} = \mbA\bth\\
\bth &=& \frac{\bbe + \mbetrue\bbe^*}{\lr{1 - \kappa}\lr{1-|\mbetrue|^2}} = \mbA^{-1}\bbe,
\end{eqnarray}
The lensing shear $\mbe_s$ is obtained from the $\mbetrue$ by some averaging process and it should be studied separately. 
As mentioned above we here only consider how to obtain $\mbetrue$ precisely.

Let ``$G(\bth)$'' and ``$\tilG(\bbe)$'' be the brightness distribution of galaxy in the 
lens and the zero plane, respectively. 
The image moments of the galaxy in the zero plane is defined as
\begin{eqnarray}
\MOM{N}{M}(\tilG) = \int d^2\beta \bbe^N_M \tilG(\bbe) \tilW\lr{{\bbe^2_0}/{\sigma_W^2}},
\end{eqnarray}
where
\begin{eqnarray}
\bbe^N_M&\equiv&\bbe^{\frac{N+M}{2}}\bbe^{*\frac{N-M}{2}},
\end{eqnarray}
and $W$ is the weight function which must be a concentric function, e.g. circular Gaussian function,  in the zero plane, and $\sigma_W$ is the weight scale.
This moment can be measured in the lens plane as 
\begin{eqnarray}
\MOM{N}{M}(G) = \int d^2\theta \lr{\mbA\bth}^N_M G(\bth) \tilW\lr{{\lr{\mbA\bth}^2_0}/{\sigma_W^2}},
\end{eqnarray}
where we ignored the scalar coefficient from Jacobian since it is not important for this study.

From the definition of the zero plane,  the dipole and the ellipticity of the image must be 0.   
Thus the complex moments satisfy the following identities. 
\begin{eqnarray}
\label{eq:Mzero_1}
\MOM{1}{1}(\tilG) &=& \int d^2\beta \bbe^1_1 \tilG(\bbe) \tilW\lr{{\bbe^2_0}/{\sigma_W^2}}
 = \int d^2\theta \lr{\mbA\bth}^1_1 G(\bth) \tilW\lr{{\lr{\mbA\bth}^2_0}/{\sigma_W^2}} = 0\\
\label{eq:Mzero_2}
\MOM{2}{2}(\tilG) &=& \int d^2\beta \bbe^2_2 \tilG(\bbe) \tilW\lr{{\bbe^2_0}/{\sigma_W^2}}
 = \int d^2\theta \lr{\mbA\bth}^2_2 G(\bth) \tilW\lr{{\lr{\mbA\bth}^2_0}/{\sigma_W^2}} = 0.
\end{eqnarray}
One can determine the zero plane by finding the centroid and the ellipticity which satisfy these equations \ref{eq:Mzero_1} and \ref{eq:Mzero_2}.

When the galaxy has the photon noise with scale $\sigma_{PN}$, the signal-to-noise ratio $\rSN$ is defined as 
\begin{eqnarray}
\rSN \equiv \frac{\MOM{0}{0}(\tilG)}{\sigma_{PN}\sqrt{\MOM{0}{0}(\tilW)}},
\end{eqnarray}
where
\begin{eqnarray}
\MOM{N}{M}(\tilW) = \int d^2\beta \bbe^N_M \tilW^2\lr{{\bbe^2_0}/{\sigma_W^2}} = \int d^2\theta \lr{\mbA\bth}^N_M \tilW^2\lr{{\lr{\mbA\bth}^2_0}/{\sigma_W^2}},
\end{eqnarray}
and the weight scale is determined to maximize the signal-to-noise ratio, so the weight scale is determined 
by  
${\partial\rSN}/{\partial\sigma_W^2} = 0$. The condition can be described as
\begin{eqnarray}
\frac{\MOM{'2}{0}(\tilG)}{\MOM{0}{0}(\tilG)}
-
\frac{\MOM{'2}{0}(\tilW)}{\MOM{0}{0}(\tilW)}
=0.
\end{eqnarray}
where $\MOM{'N}{M}(\tilG) = \int d^2\beta \bbe^N_M \tilG(\bbe) \tilW'\lr{\bbe^2_0/\sigma_W^2}$ and $\tilW'(x) = {\partial \tilW(x)}/{\partial x}$. 
If the weight function is a Gaussian function, the weight scale $\sigma_W$ and Gaussian radius $r_g$ is determined as
\begin{eqnarray}
\label{eq:g_r}
\sigma_W^2 = 2r_g^2 = 2\frac{\MOM{2}{0}(\tilG)}{\MOM{0}{0}(\tilG)}.
\end{eqnarray}

\subsection{PSF correction}
PSF correction in the framework of ERA method makes also use of the zero plane as follows. 
The observed galaxy image(smeared image) ``$S(\bth)$'' is obtained by convolution of an intrinsic image 
 $G(\bth)$ and PSF ``$P(\bth)$''
\begin{eqnarray}
S(\bth) = G(\bth)*P(\bth),
\end{eqnarray}
where $*$ represents the convolution.  In the zero plane this relation becomes
\begin{eqnarray}
\tilS(\bbe) = \tilG(\bbe)*\tilP(\bbe).
\end{eqnarray}

Let us imagine that the PSF has a concentric function in the zero plane. 
If so,  the smeared Galaxy is also concentric. 
This means that the smeared galaxy image satisfies equation \ref{eq:Mzero_2} for $S$ instead of $G$ 
and has the ellipticity $\mbetrue$ in the lens plane.
Therefore the true ellipticity $\mbetrue$ can be measured simply by measuring the shape of 
the smeared galaxy image.
However, in real analysis, the shape of PSF is not concentric in general but more complicated shape 
and moreover the shape changes in each exposure. 
ERA method creates such an ideal situation by re-smearing the smeared galaxy image again by re-smearing function(RSF) 

Let ``$\tilI(\bbe^2_0/\sigma_I^2)$'' be the idealized PSF which is a concentric function in the zero plane,
the scale $\sigma_I$ is arbitorary but it should have slightly large radius than PSF.
The Re-smeared function ``$\tilR(\bbe)$'' is then defined in the zero plane as
\begin{eqnarray}
\tilR(\bbe) = \tilI(\bbe^2_0/\sigma_I^2)\otimes\tilP(\bbe),
\end{eqnarray}
where $\otimes$ denotes the deconvolution.
Then the re-smeared galaxy ``$\tilReS(\bbe)$'' is defined as
\begin{eqnarray}
\tilReS(\bbe) = \tilS(\bbe)*\tilR(\bbe) = \tilG(\bbe)*\tilI(\bbe^2_0/\sigma_I^2),
\end{eqnarray}
and also $\tilReS(\bbe)$ satisfies equation \ref{eq:Mzero_2} as
\begin{eqnarray}
\label{eq:Mzero_2RES}
\MOM{2}{2}(\tilReS)
 &=& \int d^2\beta \bbe^2_2 \tilReS(\bbe) \tilW\lr{{\bbe^2_0}/{\sigma_W^2}}
\nonumber\\
 &=& \int d^2\beta \bbe^2_2 \lr{\tilG(\bbe)*\tilI(\bbe^2_0/\sigma_I^2)} \tilW\lr{{\bbe^2_0}/{\sigma_W^2}}
\nonumber\\
 &=& \int d^2\theta \lr{\mbA\bth}^2_2 \lr{G(\bth)*\tilI((\mbA\bth)^2_0/\sigma_I^2)} \tilW\lr{{\lr{\mbA\bth}^2_0}/{\sigma_W^2}}
\nonumber\\
 &=& \int d^2\theta \lr{\mbA\bth}^2_2 \lr{S(\bth)*\tilR(\mbA\bth)} \tilW\lr{{\lr{\mbA\bth}^2_0}/{\sigma_W^2}}
 = 0,
\end{eqnarray}
Therefore the ellipticity $\mbetrue$ which satisfies this equation is the PSF corrected ellipticity.

\section{photon noise correction}
photon noise is Poisson noise of flux from the atmosphere and thus adds random count ``$N(\bth)$'' on the image of objects.
In this paper, we assume that star images used in PSF measurement have high brightness so that we can 
ignore the photon noise effect on them. 
The observed galaxy image ``$O(\bth)$'' is smeared by PSF and has photon noise as follows
\begin{eqnarray}
\label{eq:OBS}
O(\bth) = S(\bth) + N(\bth).
\end{eqnarray}
The statistics of photon noise is described by the two point correlation function only as follows
\begin{eqnarray}
\label{eq:PRND}
\lrt{N(\bth)}&=&0\\
\lrt{N(\bth)N(\bth')}&=&\sigma^2_{PN}\delta(\bth-\bth'),
\end{eqnarray}
where the bracket means the average taken over an enough number of different random countfields. 
In the zero plane they satisfy
\begin{eqnarray}
\label{eq:PRND_zero}
\lrt{\tilN(\bbe)}&=&0\\
\lrt{\tilN(\bbe)\tilN(\bbe')}&=&\sigma^2_{PN}\delta(\bbe-\bbe'),
\end{eqnarray}
because moving to other plane is just 2-dimensional spatial scale change.

\subsection{photon noise effect for ellipticity measurement without PSF correction}
The photon noise changes the shape of a galaxy and so the ellipticity of galaxy.
Let $\mbe'$ and $\tmbe'$ be the observed ellipticity of a galaxy image with photon noise in the lens plane and in the zero plane, respectively, and 
$\Dmbe$ and $\Dtmbe$ be the additional changes of ellipticities from the true values, respectively. Thus we have  
\begin{eqnarray}
\tmbe' &=& \Dtmbe\\
\mbe'  &=& \mbetrue + \Dmbe,
\end{eqnarray}
then the relations between these ellipticities are related each other by the following equations
\begin{eqnarray}
\tmbe' &=& \frac{\mbe' - \mbetrue}{1 - \mbe'\mbetrue^*}\\
\mbe' &=& \frac{\tmbe' + \mbetrue}{1 + \tmbe'\mbetrue^*}
\end{eqnarray}
These equations lead to the fllowing relations.
\begin{eqnarray}
\label{eq:addte}
\Dtmbe &=& \frac{\Dmbe}{1 - |\mbetrue|^2 - \Dmbe\mbetrue^*}\equiv\frac{\Dmbe'}{1 - \Dmbe'\mbetrue^*}\\
\label{eq:adde}
\Dmbe' &\equiv& \frac{\Dmbe}{1 - |\mbetrue|^2} = \frac{\Dtmbe}{1 + \Dtmbe\mbetrue^*}.
\end{eqnarray}

Since the photon noise changes ellipticity of galaxy, the corresponding zero plane(we call it as the zero plane with noise) is not the original one but is determined as follows
\begin{eqnarray}
\label{eq:shape_lens}
&&\int d^2\beta' \lr{\bbe'}^2_2 \lr{\tilG(\bbe)+\tilN(\bbe)} \tilW((\bbe')^2_0/\sigma_W^2) = 
\nonumber\\
&&\int d^2\theta \lr{\mbA'\bth}^2_2 \lr{G(\bth)+N(\bth)} \tilW((\mbA'\bth)^2_0/\sigma_W^2) = 0,
\end{eqnarray}
where $\bbe' = \mbA'\bth$ is the coordinates in the zero plane with noise.
The coordinate can be written in terms of the observed ellipticity as $\mbA'\bth = \lr{1 - \kappa}\lr{\bth - \mbe'\bth^*}$, then it is related with $\bbe$ as
\begin{eqnarray}
\bbe' = \mbA'\bth  = \frac{\bbe-\Dtmbe\bbe^*}{1+\Dtmbe\mbetrue^*} = \bbe - \frac{\Dtmbe}{1+\Dtmbe\mbetrue^*} \lr{\bbe^*+\mbetrue^*\bbe},
\end{eqnarray}
Thus the equation \ref{eq:shape_lens} can be written in the original zero plane as follows
\begin{eqnarray}
\label{eq:shape_zero}
\int d^2\beta \lr{\frac{\bbe-\Dtmbe\bbe^*}{1+\Dtmbe\mbetrue^*}}^2_2 \lr{\tilG(\bbe)+\tilN(\bbe)} W\lr{\lr{\frac{\bbe-\Dtmbe\bbe^*}{1+\Dtmbe\mbetrue^*}}^2_0/\sigma_W^2} = 0.
\end{eqnarray}
To derive the equations for the first order and the second order changes of the ellipticity we 
expand the weight function with respect to $\Dtmbe$ up to the second order.
\begin{eqnarray}
\label{eq:expW}
W\lr{\lr{\frac{\bbe-\Dtmbe\bbe^*}{1+\Dtmbe\mbetrue^*}}^2_0/\sigma_W^2} 
&\approx&
W(\bbe^2_0/\sigma_W^2) 
\nonumber\\&&\hspace{-150pt}
-\frac{W'(\bbe^2_0/\sigma_W^2)}{\sigma^2_W}\lr{\lr{\lr{2\Dtmbe\cdot\mbetrue - \lr{1-|\mbetrue|^2}|\Dtmbe|^2-4\lr{\Dtmbe\cdot\mbetrue}^2}}\bbe^2_0
+2\lr{1-2\Dtmbe\cdot\mbetrue}\Dtmbe\cdot\bbe^2_2}
\nonumber\\&&
+\frac{2W''(\bbe^2_0/\sigma_W^2) }{\sigma_W^4}\lr{\Dtmbe\cdot\mbetrue\bbe^2_0+\Dtmbe\cdot\bbe^2_2}^2
\nonumber\\&&\hspace{-0pt}\approx
W(\bbe^2_0/\sigma_W^2) 
-\frac{2W'(\bbe^2_0/\sigma_W^2)}{\sigma^2_W}\lr{\lr{\Dtmbe\cdot\mbetrue}\lr{\bbe^2_0-\Dtmbe\bbe^2_{-2}}
+\Dtmbe\cdot\bbe^2_2}
\nonumber\\&&\hspace{-0pt}
+\frac{2\Dtmbe W''(\bbe^2_0/\sigma_W^2) }{\sigma_W^4}\lr{\Dtmbe\cdot\mbetrue}\bbe^4_{-2}
\end{eqnarray}
where we neglected terms which do not contribute in the moment in 
the further calculations and used the fact 
$2\Dtmbe\cdot\mbe = \Dtmbe^*\mbe + \Dtmbe\mbe^*$.
Thus the equation \ref{eq:shape_zero} can be approximated up to the second order in $\Dtmbe$ as
\begin{eqnarray}
\lrs{\Dtmbe\lr{-2-\frac{\SHP{'4}{0}}{\sigma_W^2}}+|\Dtmbe_{(1)}|^2\frac{\mbetrue}{\sigma_W^2}\lr{3\SHP{'4}{0}+\frac{\SHP{''6}{0}}{\sigma_W^2}}}_{(G)} + \lrs{\SHP{2}{2}-\frac{\mbetrue\Dtmbe^*\SHP{'4}{2}}{\sigma_W^2}}_{(N)} = 0,
\end{eqnarray}
where the moments are normalized by the quadrupole moment of $\tilG$,
so $\SHP{N}{M}(A) \equiv \MOM{N}{M}(A)/\MOM{2}{0}(G)$, 
By separating the 1st order terms and the 2nd order terms, we obtain the following equations
\begin{eqnarray}
\label{eq:shape_zero1st}
\Dtmbe_{(1)} &=& \frac{1}{2+\frac{\SHP{'4}{0}(G)}{\sigma_W^2}}\SHP{2}{2}(N) \equiv \CoefW\SHP{2}{2}(N)  \\
\label{eq:shape_zero2nd}
\Dtmbe_{(2)} &=& \CoefW\mbetrue\lr{|\Dtmbe_{(1)}|^2\lr{\frac{3\SHP{'4}{0}(G)}{\sigma_W^2}+\frac{\SHP{''6}{0}(G)}{\sigma_W^4}}-\Dtmbe_{(1)}^*\frac{\SHP{'4}{2}(N)}{\sigma_W^2}}.
\end{eqnarray}
The statistical averages of the 1st order and 2nd order additional ellipticity are thus  
\begin{eqnarray}
\lrt{\Dtmbe_{(1)}} &=& 0\\
\lrt{|\Dtmbe_{(1)}|^2} &=& \CoefW^2\lrt{\SHP{2}{2}(N)\SHP{2}{-2}(N)} =
\lr{\CoefW\frac{\SHP{0}{0}(G)}{\rSN}}^2\frac{\MOM{4}{0}(W)}{\MOM{0}{0}(W)}
 \eqG \frac{2}{\rSN^2}\\
\lrt{\Dtmbe_{(1)}^2} &=& 0\\
\label{eq:tildePNcor2}
\lrt{\Dtmbe_{(2)}} &=& 
\mbetrue\lr{\CoefW\frac{\SHP{0}{0}(G)}{\rSN}}^2
\Biggl(\CoefW\frac{\MOM{4}{0}(W)}{\MOM{0}{0}(W)}\lr{\frac{3\SHP{'4}{0}(G)}{\sigma_W^2}+\frac{\SHP{''6}{0}(G)}{\sigma_W^4}}
\nonumber\\&&-
\frac{1}{\sigma_W^2}\frac{\MOM{'6}{0}(W)}{\MOM{0}{0}(W)}\Biggr) \eqG 0.
\end{eqnarray}
where $\eqG$ means the analytical result in the case that the galaxy image and weight function are both elliptical Gaussian, and in this situation, $\CoefW\eqG1$, and
$\lrt{}$ means the ensemble  average over enough number of different random count images,
\begin{eqnarray}
\lrt{\SHP{N}{M}(N)\SHP{O}{P}(N)}& = &
\lr{\frac{\sigma_{PN}^2}{\MOM{2}{0}(G)}}^2\lrt{\MOM{N}{M}(N)\MOM{O}{P}(N)}
\nonumber\\& = &
\lr{\frac{\SHP{0}{0}(G)}{\rSN}}^2\frac{\MOM{N+O}{M+P}(W)}{\MOM{0}{0}(W)}.
\end{eqnarray}
This result means that the photon noise almost does not make a systematic error and makes only concentric dispersion in the zero planes, so it behaves like intrinsic ellipticity noise.

The additional ellipticity in the lens plane is obtained as 
\begin{eqnarray}
\lrt{\Dmbe} = \lr{1-|\mbe_g|^2}\lrt{\frac{\Dtmbe}{1+\Dtmbe\mbetrue^*}}
\approx
\lr{1-|\mbe_g|^2}\lrt{\Dtmbe_{(2)}}
\equiv
\Dmbe_{cor}
\end{eqnarray}

Finally, we can obtain photon noise corrected ellipticity $\mbe_{cor}$ as follows.
\begin{eqnarray}
\mbe_{cor} = \mbe' - \Dmbe_{cor},
\end{eqnarray}
and $\mbe_s$ is obtained as ellipticity which satisfying 
\begin{eqnarray}
\lrt{\frac{\mbe_{cor} - \mbe_s}{1 - \mbe_{cor}\mbe_s^*}}=0.
\end{eqnarray}
where $\lrt{}$ means the ensemble average over enough number of different backgrond galaxy images.

Even the galaxy image is an elliptical Gaussian or similar shape which has $\lrt{\Dtmbe_{(2)}}=0$, the nonlinear average is needed.
This is because the photon noise makes only concentric distribution in ellipticity space, but the distribution is not concentric in the image plane.
Thus if one simply takes a linear average of $\mbe_{cor} $, it suffers from noise bias.

Although the 2nd order is close to 0, the increase of ellipticity dispersion by the 1st order effect is important, e.g. when we use the weighted average for measured galaxies to obtain shear by taking the average.
As simple example,
we can define the weight of galaxies when measuring shear by averaging as
\begin{eqnarray}
w = \frac{\sigma_{\rm int}^2}{\sigma_{\rm int}^2 + \lrt{|\Dtmbe_{(1)}|^2}}
\end{eqnarray}
where $\sigma_{\rm int}$ is the standard deviation of the intrinsic ellipticity
which is defined as 
\begin{eqnarray}
\sigma_{\rm int} = \sqrt{\lrt{\mbe_z\mbe_z^*}}
\end{eqnarray}

\subsection{photon noise effect for ellipticity measurement with PSF correction}
ERA method re-smears the galaxy image again by re-smearing function to reshape PSF to have an idealized PSF. 
The idealized PSF has the same ellipticity with  PSF corrected ellipticity which is affected by photon noise, so the equation \ref{eq:Mzero_2RES} changes by photon noise as follows
\begin{eqnarray}
\label{eq:Mzero_2RESN}
\MOM{2}{2}(\tilReS)
 &=& \int d^2\theta \lr{\mbA'\bth}^2_2 \lr{S(\bth) + N(\bth)}*R(\mbA'\bth) W\lr{{\lr{\mbA'\bth}^2_0}/{\sigma_W^2}}
 = 0,
\end{eqnarray}
where $R(\mbA'\bth) = I(\lr{\mbA'\bth}^2_0/\sigma_W^2)\otimes P(\bth)$ is the re-smearing function. 

In this case, we expand the re-smearing function up to the second order in  
  $\Dtmbe$ instead of the weight function in the equation \ref{eq:expW}. 
Then  \ref{eq:Mzero_2RESN} gives the following equations for the first order and the second order chnage in the ellipticity. 
\begin{eqnarray}
\label{eq:shear_zero1st}
\Dtmbe_{(1)} &=& \frac{\SHP{2}{2}(\ReN)}{2+\frac{\SHP{'4}{0}(\ReS)}{\sigma_W^2}+\frac{\SHP{2}{2}(\ReS^{'2}_{-2})}{\sigma_R^2}} \equiv \CoefP\SHP{2}{2}(\ReN)  \\
\label{eq:shear_zero2nd}
\Dtmbe_{(2)} &=& 
\CoefP\Dtmbe_{(1)}\lr{2\Dtmbe_{(1)}\cdot\mbetrue}
\Biggl(
\frac{3\SHP{'4}{0}(\ReS)}{\sigma_W^2}
+\frac{\SHP{''6}{0}(\ReS)}{\sigma_W^4}
+\frac{2\SHP{2}{0}(\ReS^{'2}_{0})
+\SHP{2}{2}(\ReS^{'2}_{-2})}{\sigma_R^2}
\nonumber\\&&\hspace{150pt}
+\frac{\SHP{2}{2}(\ReS^{''4}_{-2})}{\sigma_R^4}
+\frac{\SHP{4}{0}(\ReS^{2}_{0})+\SHP{4}{2}(\ReS^{2}_{-2})}{\sigma_R^2\sigma_W^2}
\Biggr)
\nonumber\\&&\hspace{-50pt}-
\CoefP\Biggl(
\Dtmbe_{(1)}\lr{2\SHP{2}{0}(\ReN)
+\frac{\SHP{'4}{0}(\ReN)+\mbetrue^*\SHP{'4}{2}(\ReN)}{\sigma_W^2}
+\frac{\SHP{2}{2}(\ReN^{'2}_{-2})+\mbetrue^*\SHP{2}{2}(\ReN^{'2}_{0})}{\sigma_R^2}
}
\nonumber\\&&\hspace{0pt}
+\Dtmbe^*_{(1)}\lr{
\frac{\SHP{'4}{4}(\ReN)+\mbetrue\SHP{'4}{2}(\ReN)}{\sigma_W^2}
+\frac{\SHP{2}{2}(\ReN^{'2}_{2})+\mbetrue\SHP{2}{2}(\ReN^{'2}_{0})}{\sigma_R^2}
}\Biggr).
\end{eqnarray}
where, ${{\tilde S}^{ReN}_{M}} \equiv \tilS*(\tilR\beta^{N}_{M})$ and similarly for ${{\tilde N}^{ReN}_{M}}$,
and the shape $\SHP{N}{M}$ is normalized by quadrupole moments of $\tilReS$.
$\Dtmbe_{(1)}$ is the statistical noise and $\lrt{\Dtmbe_{(2)}}$ is the noise bias in PSF corrected ellipticity
The dispersion and bias from the photon noise can be predicted by taking the average of equation \ref{eq:shear_zero1st} and \ref{eq:shear_zero2nd}. 
\begin{eqnarray}
\lrt{|\Dmbe|^2} &\approx& \lrt{|\Dmbe_{(1)}|^2}  = \lr{1-|\mbetrue|^2}^2\lrt{|\Dtmbe_{(1)}|^2}
\nonumber\\&=&
\CoefP^2\sigma^2_{PN}\lr{1-|\mbetrue|^2}^2{\RWRW{\tilR*W^2_2, \tilR*W^2_{-2}}}
\\
\lrt{\Dmbe} &\approx& \lrt{\Dmbe_{(2)}}
 = \lr{1-|\mbetrue|^2}\lrt{\Dtmbe_{(2)}}
\nonumber\\&=&
\CoefP^2\sigma^2_{PN}\lr{1-|\mbetrue|^2}
\Biggl(\CoefP\RWRW{\tilR*W^2_2, \tilR*W^2_{-2}\mbetrue + \tilR*W^2_2\mbetrue^*}
\times\nonumber\\&&\hspace{-50pt}
\Biggl(
\frac{3\SHP{'4}{0}(\ReS)}{\sigma_W^2}
+\frac{\SHP{''6}{0}(\ReS)}{\sigma_W^4}
+\frac{2\SHP{2}{0}(\ReS^{'2}_{0})
+\SHP{2}{2}(\ReS^{'2}_{-2})}{\sigma_R^2}
+\frac{\SHP{2}{2}(\ReS^{''4}_{-2})}{\sigma_R^4}
+\frac{\SHP{4}{0}(\ReS^{2}_{0})+\SHP{4}{2}(\ReS^{2}_{-2})}{\sigma_R^2\sigma_W^2}
\Biggr)
\nonumber\\&&\hspace{-75pt}-
\RWRW{\tilR*W^2_2, 2\tilR*W^2_0
+\frac{\tilR*W^{`4}_0 + \tilR*W^{`4}_2\mbetrue^*}{\sigma_W^2}
+\frac{\tilR^{`2}_{-2}*W^2_0 + \tilR^{`2}_0*W^2_2\mbetrue^*}{\sigma_R^2}
}
\nonumber\\&&\hspace{-75pt}-
\RWRW{\tilR*W^2_{-2}, \frac{
\tilR*W^{`4}_4 + \tilR*W^{`4}_2\mbetrue}{\sigma_W^2}+
\frac{\tilR^{`2}_2*W^2_2 + \tilR^{`2}_0*W^2_2\mbetrue}{\sigma_R^2}}
\Biggr)
\end{eqnarray}
where
\begin{eqnarray}
\RWRW{\tilR^N_M*W^O_P, \tilR^Q_R*W^S_T}&\equiv&\frac{\int d^2\beta \lr{\tilR^N_M*W^O_P}\lr{\tilR^Q_R*W^S_T}}{\lr{\MOM{2}{0}\lr{\ReS}}^2}\\
\tilR^N_M &\equiv& \lr{I(\bbe^2_0/\sigma^2_I)\bbe^N_M}\otimes\tilP(\bbe)\\
W^N_M &\equiv&W\lr{\bbe^2_0/\sigma^2_W}\bbe^N_M
\end{eqnarray}
Thus photon noise + PSF corrected ellipticity is obtained by the above expression for 
$\lrt{\Dmbe}$.
\section{Simulation tests using simple models of galaxy  and real galaxy images}

We test the correction method derived in the previous section by simulation using some simple models of galaxy and PSF, as well as real galaxy images in GalSim(Rowe et al. 2015) taken by HST(Hubble Space Telescope).  

As  simple models of the galaxy we use Gaussian and Sersic profiles for galaxy whose radius determined by equation \ref{eq:g_r} is 2.0 pixels and the ellipticity is 0.2.
For PSF we use Gaussian PSF, Gaussian weight function and Gaussian idealized PSF. ,
The radius of PSF is selected from [1.5, 2.0, 2.5] pixels and ellipticity from [0.0, -0.1, 0.1i].

The steps of the simulation are as follows. 
First, we create galaxy and PSF image and make the smeared galaxy image by convolving the galaxy and the PSF image. Then we add random count image for the smeared galaxy image. 
Finally, we measure ``PSF corrected ellipticity'' and ``PSF and photon noise corrected ellipticity''.  
We measure the two ellipticities with 40,000 different random count images, then we calculate the mean of the ellipticities with two times 5$\sigma$ outlier clipping.

\begin{figure*}[tbp]
\centering
\resizebox{1.0\hsize}{!}{\includegraphics{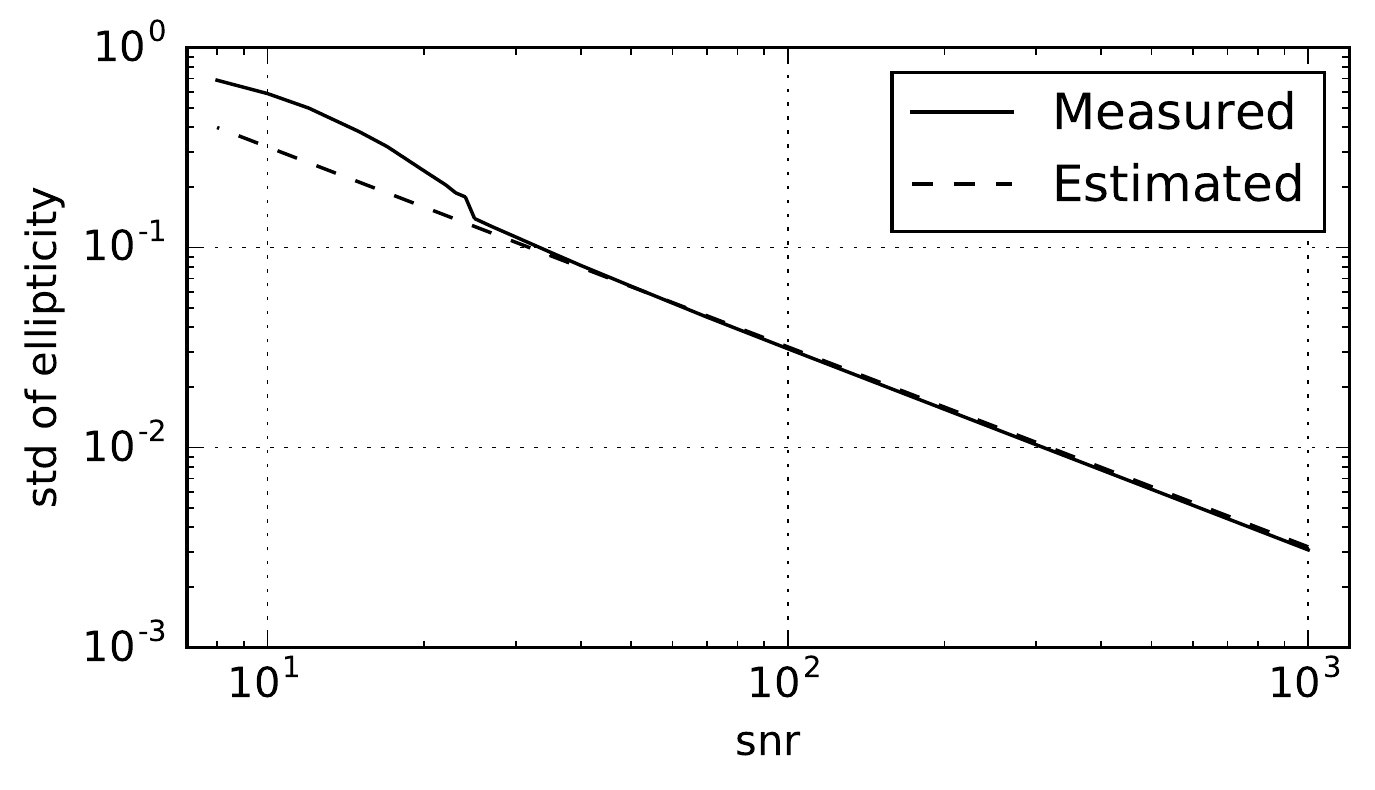}}
\caption{
\label{fig:sed_de}
Measured and estimated standard deviation of ellipticity as a function of SNR.
The horizontal and the vertical axis are SNR of the galaxy and standard deviation of ellipticity, respectively.
The solid line and dashed line mean the measured and the predicted standard deviation, respectively. 
This is one of the results of the simulations and in this figure galaxy profile is Gaussian, PSF radius = 2.0 pixels and PSF ellipticity = 0.0.
}
\end{figure*}

Figure \ref{fig:sed_de} shows the measured and the estimated standard deviation of the ellipticity as a function of SNR.
We can see that the standard deviation can be predicted quite well especially in high SNR region.
We performed other simulations using several different parameters, and the results are similar 
with the one presented in this Figure.

Figure \ref{fig:mean_de_G_20} to \ref{fig:mean_de_S_01} show the multiplicative bias of PSF corrected ellipticity and PSF and photon noise corrected ellipticity from true ellipticity as a function of SNR, where the vertical axis is the multiplicative bias $m$ defined as $m = |(\lrt{\mbe} - \mbetrue)/\mbetrue|$ where $\lrt{\mbe}$ is the mean value of ellipticity with or without photon noise correction and $\mbetrue$ is the true ellipticity.
The results show that the photon noise makes overestimation in shear analysis and it reaches $1\%$ around $\rSN = $[50,70,100] for PSF radius = [1.5, 2.0, 2.5] pixels and galaxy radius = 2.0 pixels,
but the photon noise correction can correct $80 \sim 90\%$ of multiplicative bias especially in high SNR region, the  ``high'' region here means the region where the multiplicative bias obeys the inverse square law with respect to SNR.
Figure \ref{fig:SNRborderG} and \ref{fig:SNRborderS} show boundaries of $1\%$ multiplicative bias in SNR with and without photon noise correction.

Figure \ref{fig:reducednumberG} and \ref{fig:reducednumberS} show the number ratio of galaxies we may able to use for measuring the multiplicative bias as a function of SNR. 
In the low SNR region, some of the images are rejected by some reasons such as the determined radius is too small and so on, and in this simulation, we used 5$\sigma$ clipping for rejecting outlier due to the divergence of the correction value, so the lower SNR galaxies have the higher rejected number.
The rejection number depends on the analysis parameters, so this figure shows just a sample in this simulation, but it is useful to know the typical number ratio of the galaxy we can use in real analysis.
We can see $80\% \sim 90\%$ of galaxies can be used at $\rSN = 20$.
The divergence in the correction comes from the divergence of $\CoefP$, so it may be modified by 
adding certain value in $\CoefP$ to avoiding the divergence.
We will study the modification in detail in future works.

\begin{figure*}[tbp]
\centering
\resizebox{1.0\hsize}{!}{\includegraphics{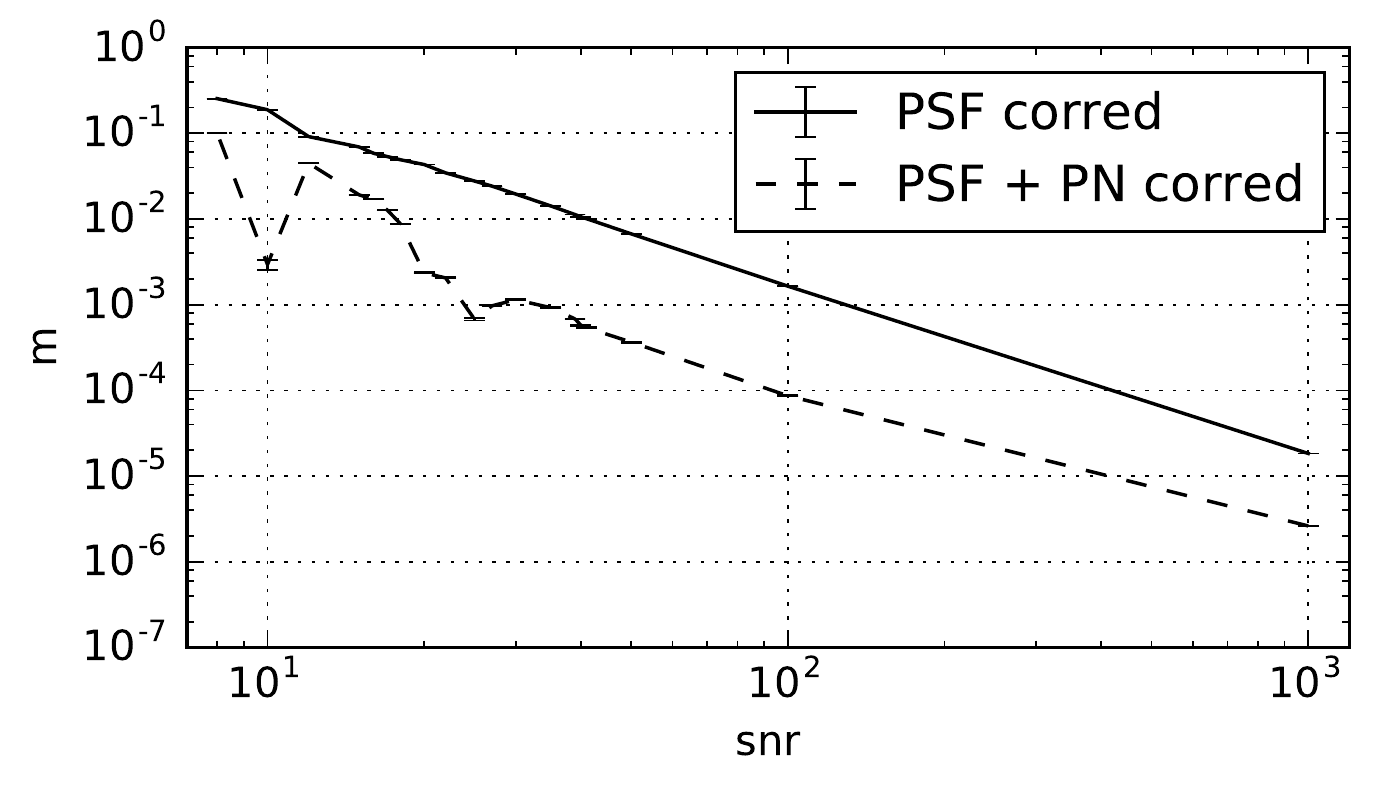}}
\caption{
\label{fig:mean_de_G_15}
The multiplicative bias of ellipticity without(solid line) and with(dashed line) the photon noise correction.
In this simulation galaxy profile is Gaussian, PSF radius is 1.5 pixels and PSF ellipticity is 0.0.
The horizontal and the vertical axis are SNR of the galaxy and multiplicative bias, respectively.
}
\end{figure*}
\begin{figure*}[tbp]
\centering
\resizebox{1.0\hsize}{!}{\includegraphics{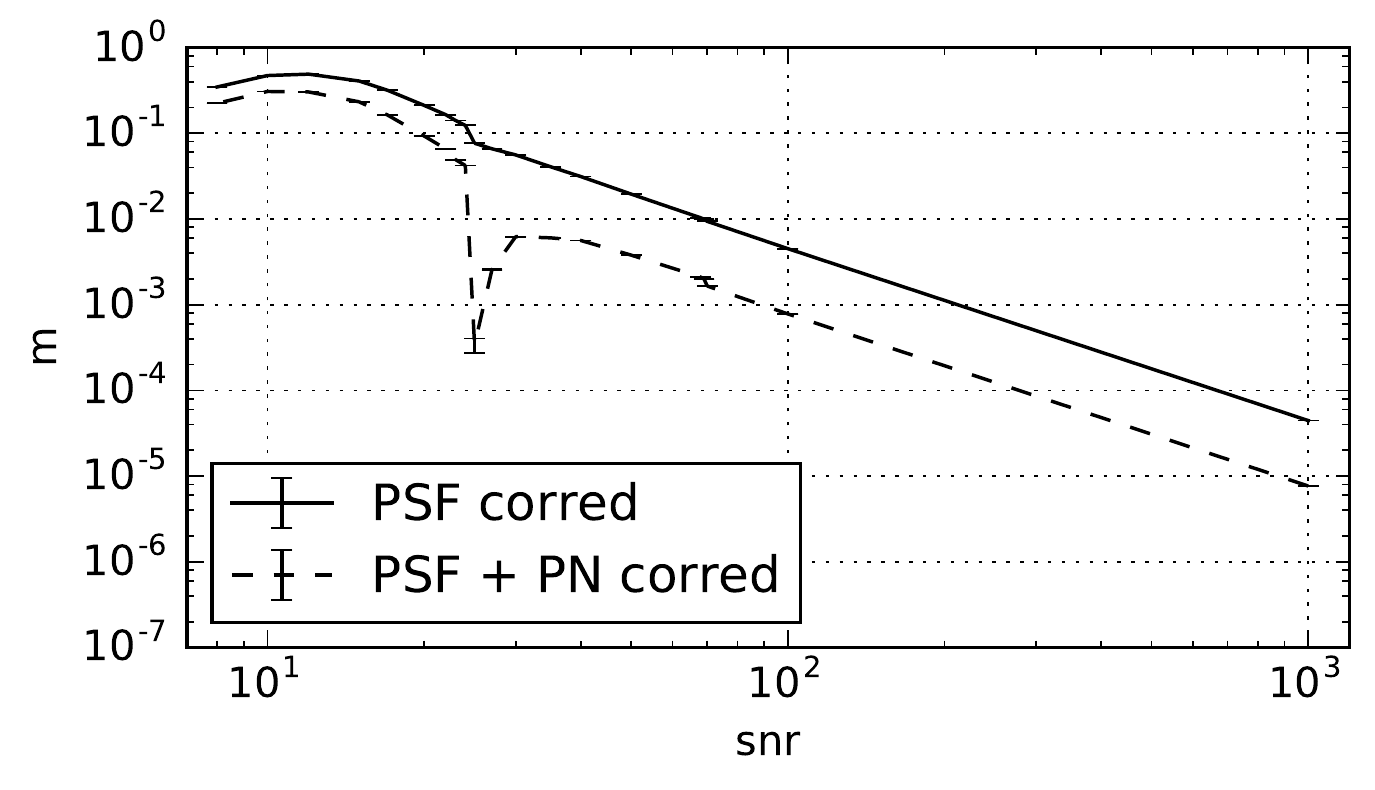}}
\caption{
\label{fig:mean_de_G_20}
Same figure as figure \ref{fig:mean_de_G_15}, except PSF radius = 2.0 pixel.
}
\end{figure*}
\begin{figure*}[tbp]
\centering
\resizebox{1.0\hsize}{!}{\includegraphics{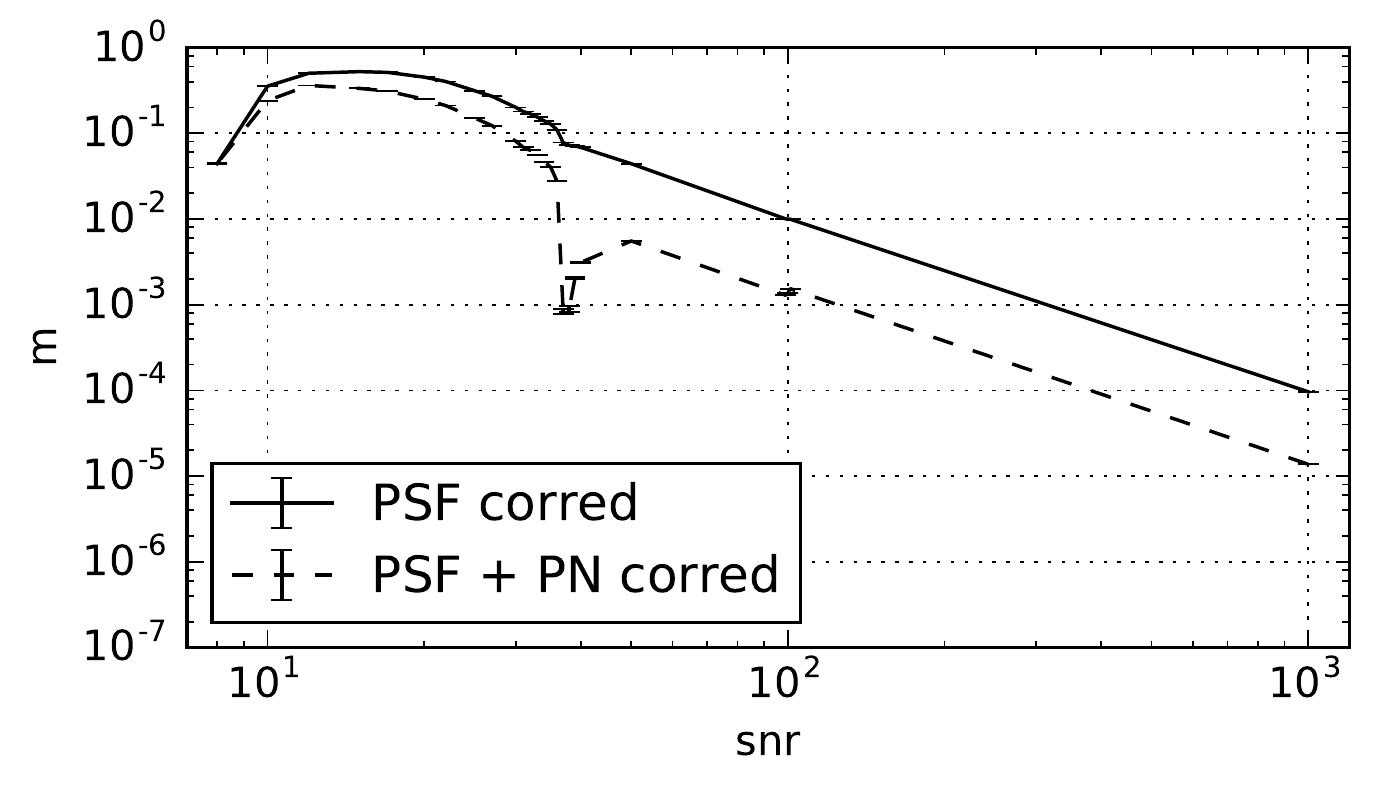}}
\caption{
\label{fig:mean_de_G_25}
Same figure as figure \ref{fig:mean_de_G_15}, except PSF radius = 2.5 pixel.
}
\end{figure*}
\begin{figure*}[tbp]
\centering
\resizebox{1.0\hsize}{!}{\includegraphics{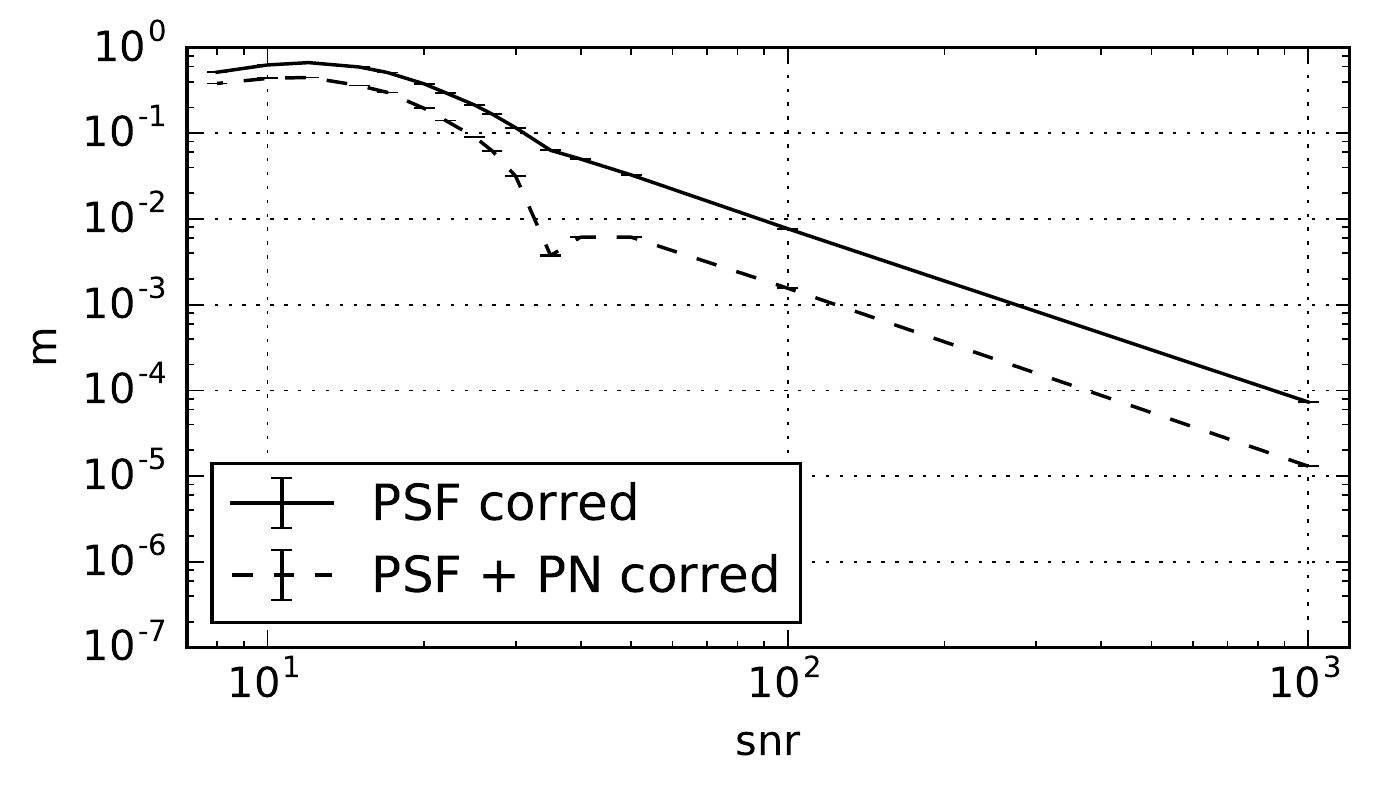}}
\caption{
\label{fig:mean_de_G_10}
Same figure as figure \ref{fig:mean_de_G_20}, except PSF ellipticity = -0.1.
}
\end{figure*}
\begin{figure*}[tbp]
\centering
\resizebox{1.0\hsize}{!}{\includegraphics{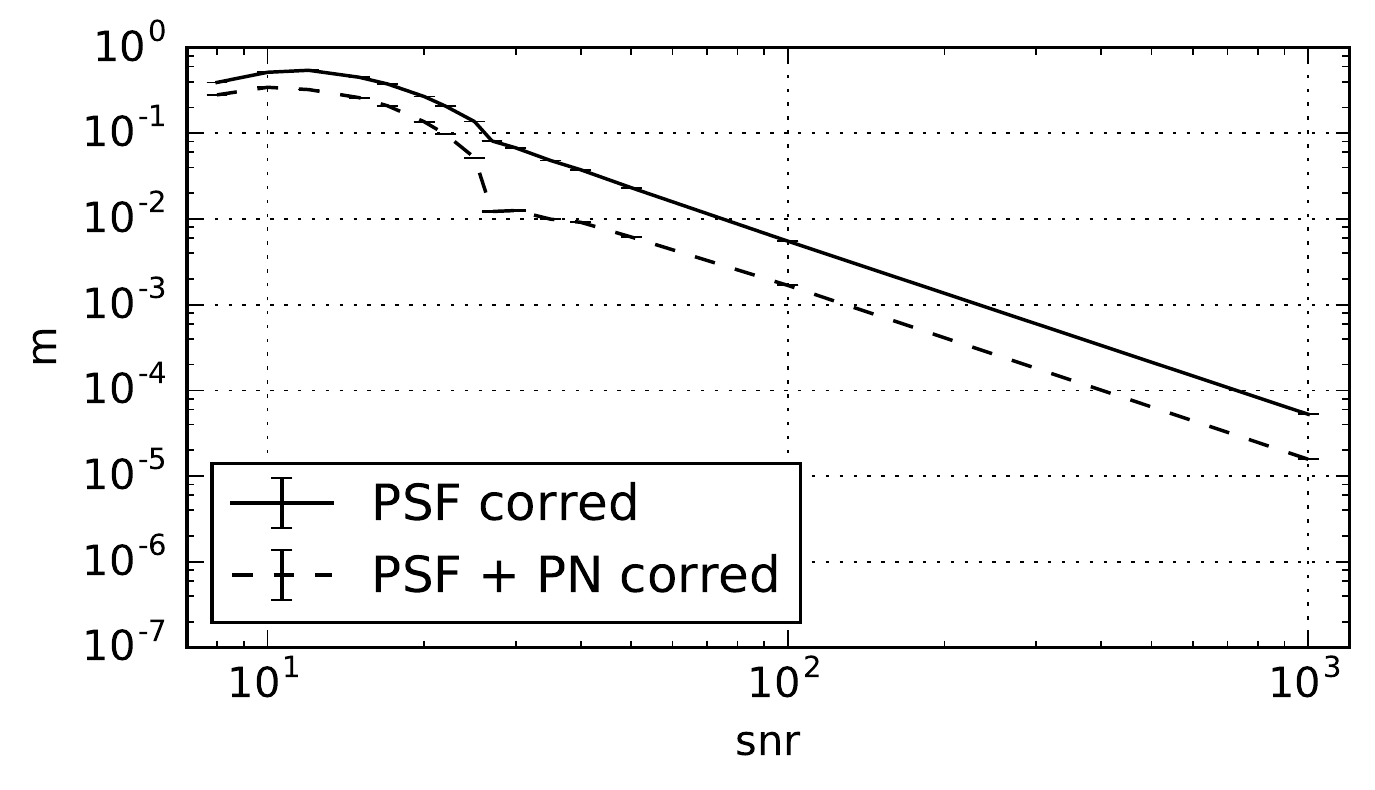}}
\caption{
\label{fig:mean_de_G_01}
Same figure as figure \ref{fig:mean_de_G_20}, except PSF ellipticity = 0.1i.
}
\end{figure*}

\begin{figure*}[tbp]
\centering
\resizebox{1.0\hsize}{!}{\includegraphics{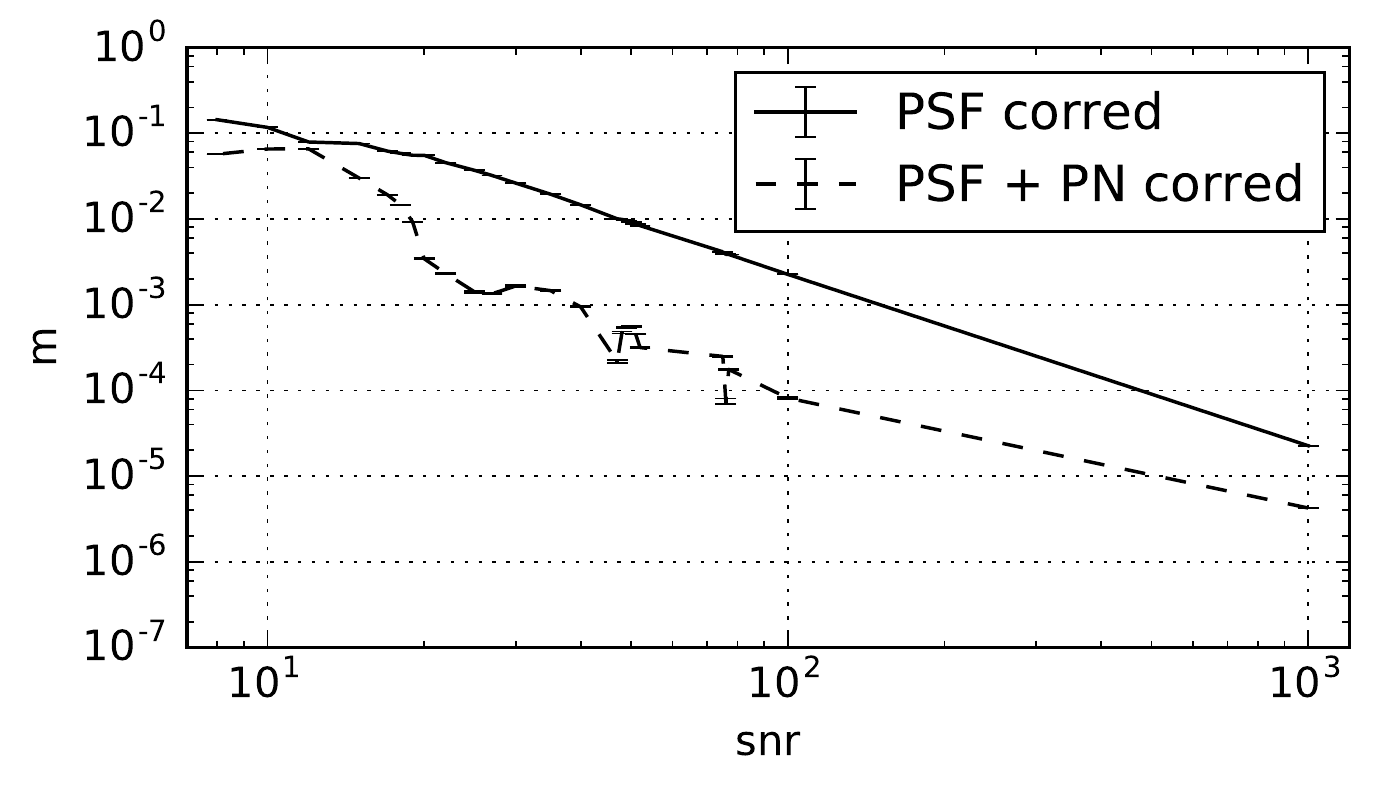}}
\caption{
\label{fig:mean_de_S_15}
Same figure as figure \ref{fig:mean_de_G_15}, except galaxy profile is Sersic.
}
\end{figure*}
\begin{figure*}[tbp]
\centering
\resizebox{1.0\hsize}{!}{\includegraphics{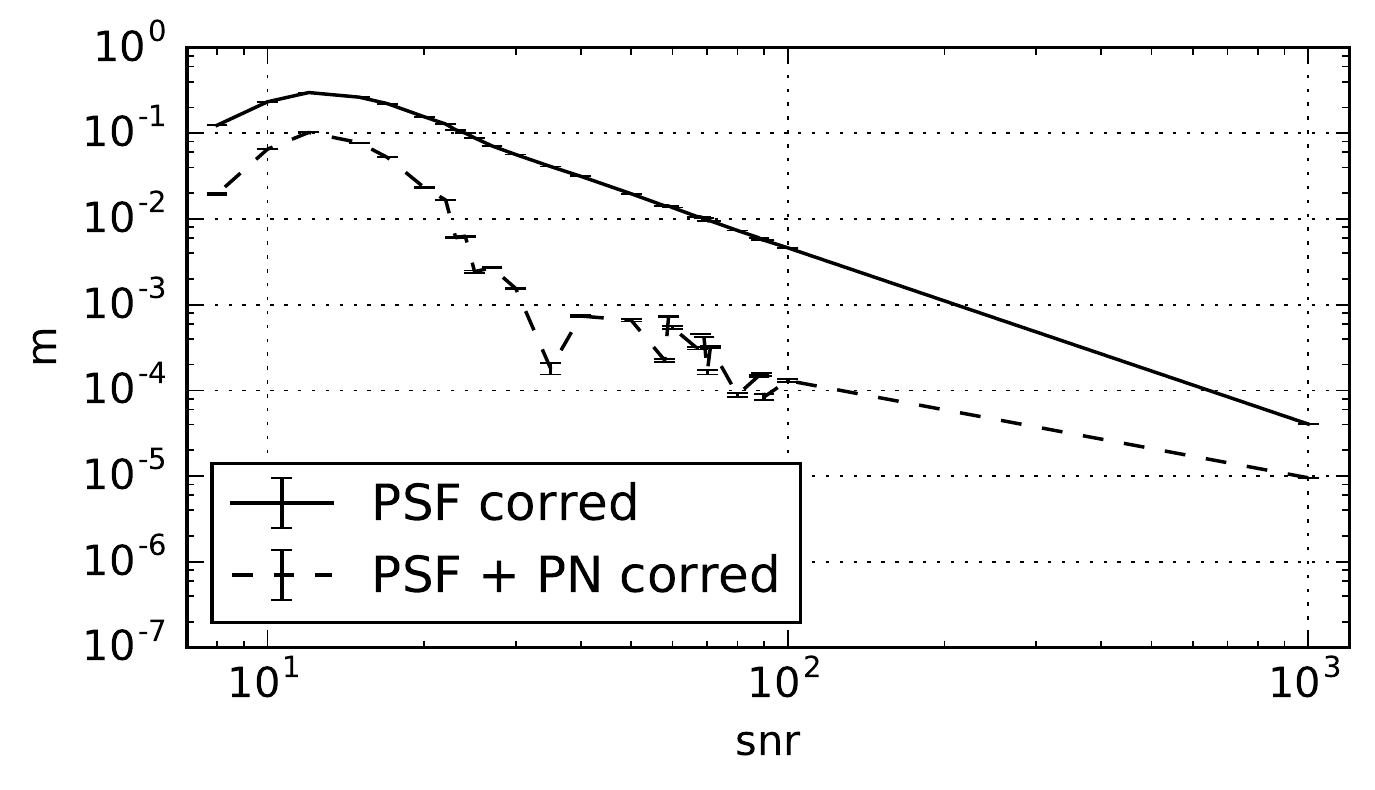}}
\caption{
\label{fig:mean_de_S_20}
Same figure as figure \ref{fig:mean_de_S_15}, except PSF radius = 2.0 pixel.
}
\end{figure*}
\begin{figure*}[tbp]
\centering
\resizebox{1.0\hsize}{!}{\includegraphics{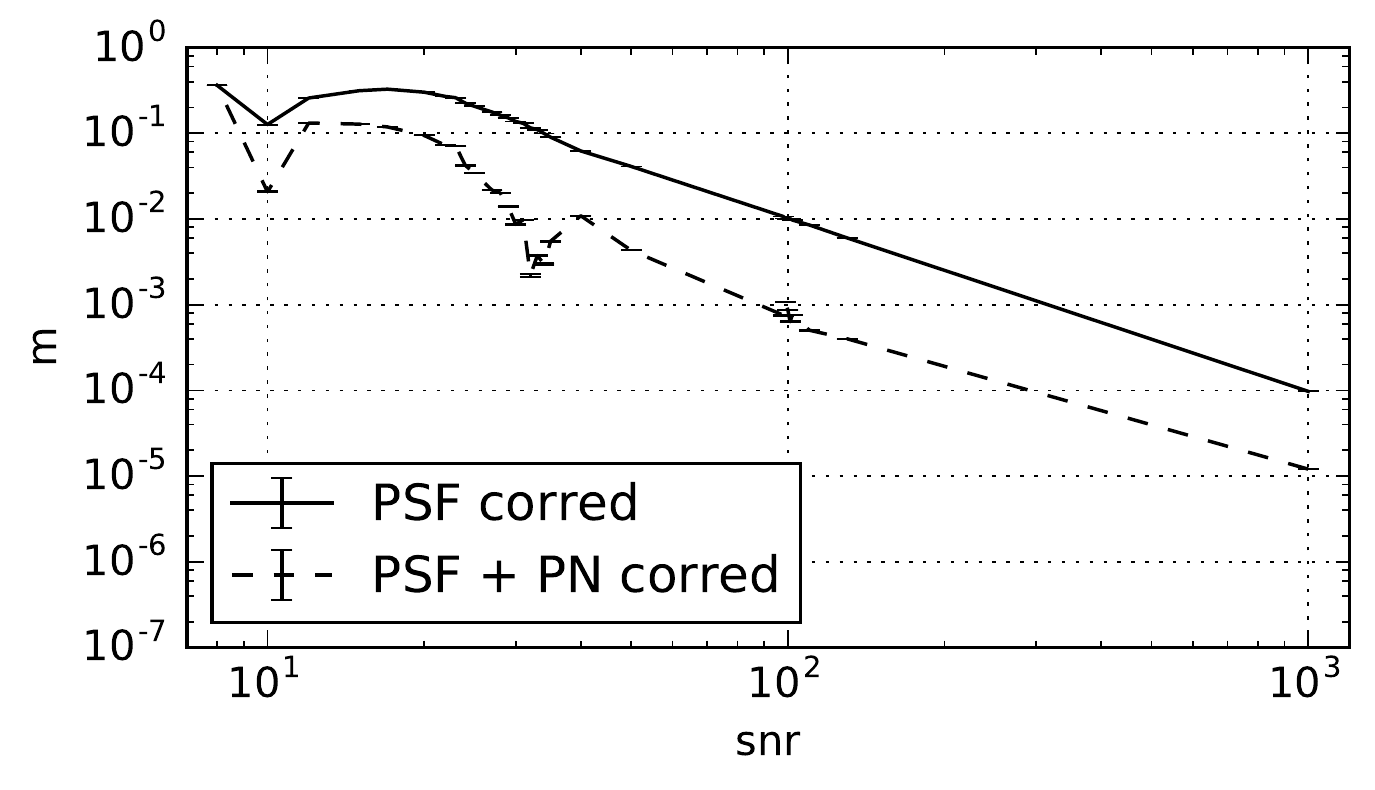}}
\caption{
\label{fig:mean_de_S_25}
Same figure as figure \ref{fig:mean_de_S_15}, except PSF radius = 2.5 pixel.
}
\end{figure*}
\begin{figure*}[tbp]
\centering
\resizebox{1.0\hsize}{!}{\includegraphics{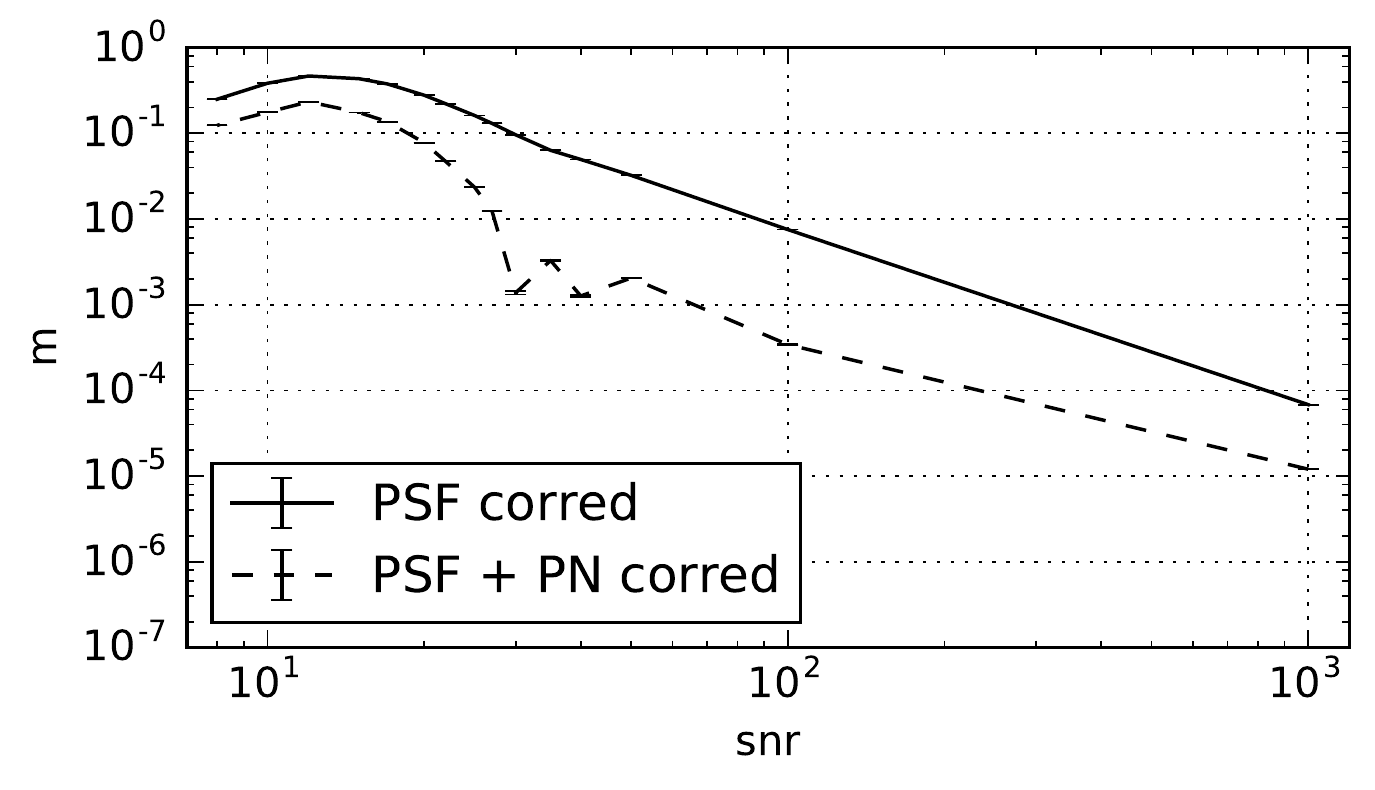}}
\caption{
\label{fig:mean_de_S_10}
Same figure as figure \ref{fig:mean_de_S_20}, except PSF ellipticity = -0.1.
}
\end{figure*}
\begin{figure*}[tbp]
\centering
\resizebox{1.0\hsize}{!}{\includegraphics{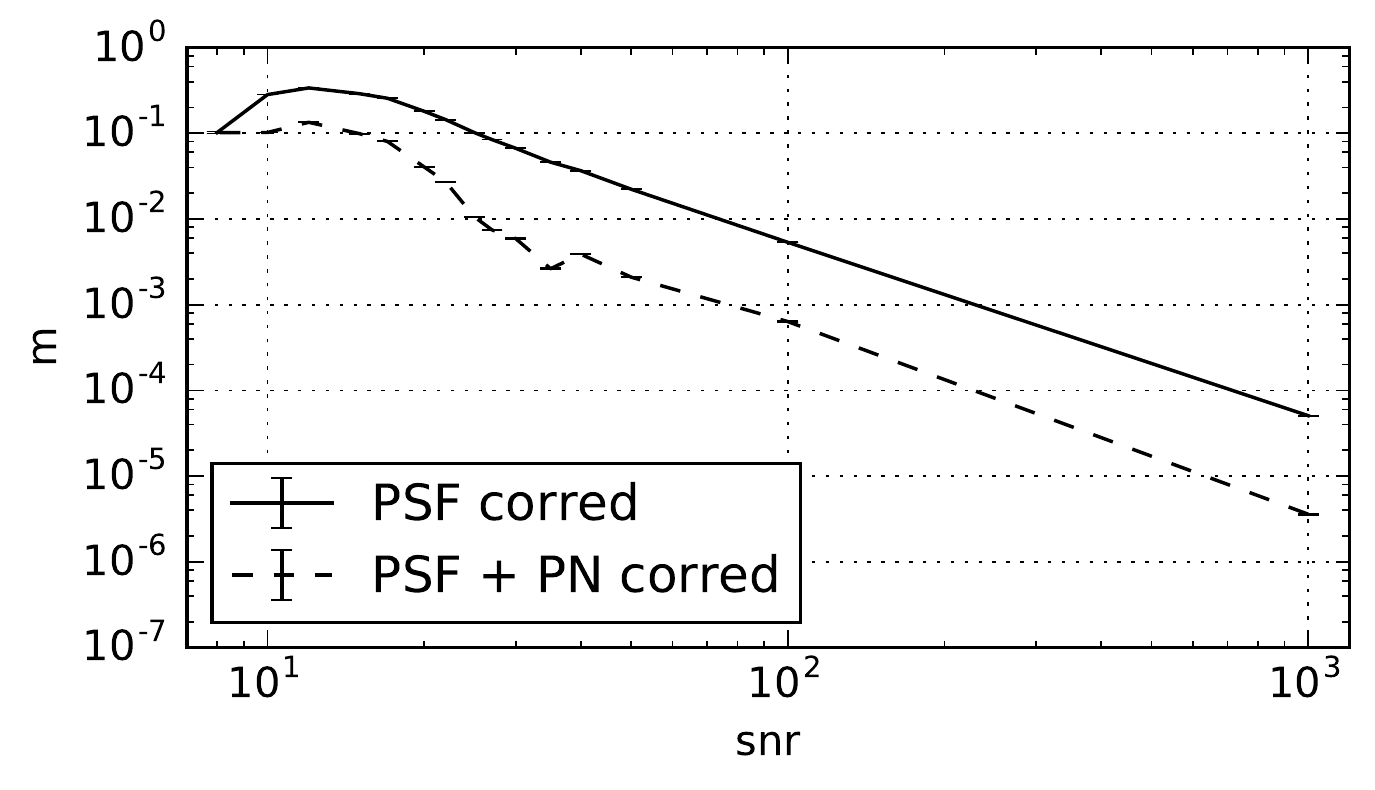}}
\caption{
\label{fig:mean_de_S_01}
Same figure as figure \ref{fig:mean_de_S_20}, except PSF ellipticity = 0.1i.
}
\end{figure*}

\begin{figure*}[tbp]
\centering
\resizebox{1.0\hsize}{!}{\includegraphics{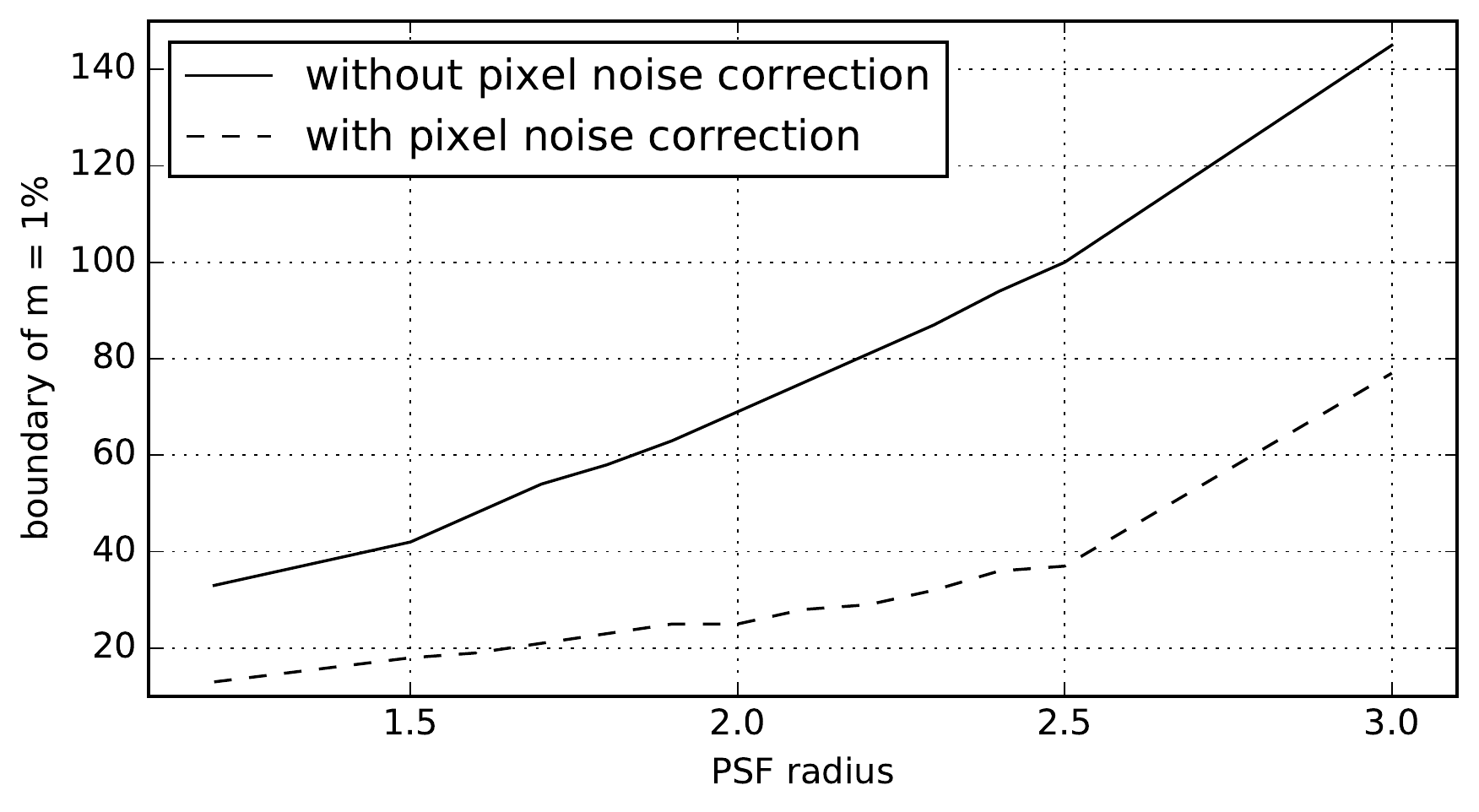}}
\caption{
\label{fig:SNRborderG}
SNR border of $1\%$ multiplicative bias with and without the photon noise correction for different PSF sizes.
The horizontal and vertical axes are PSF radius and SNR of the boundary, respectively.
The parameters used for this plot are that galaxy profile is Gaussian and PSF radius is 2.0 pixel, and 
PSF ellipticity = 0.0.
}
\end{figure*}
\begin{figure*}[tbp]
\centering
\resizebox{1.0\hsize}{!}{\includegraphics{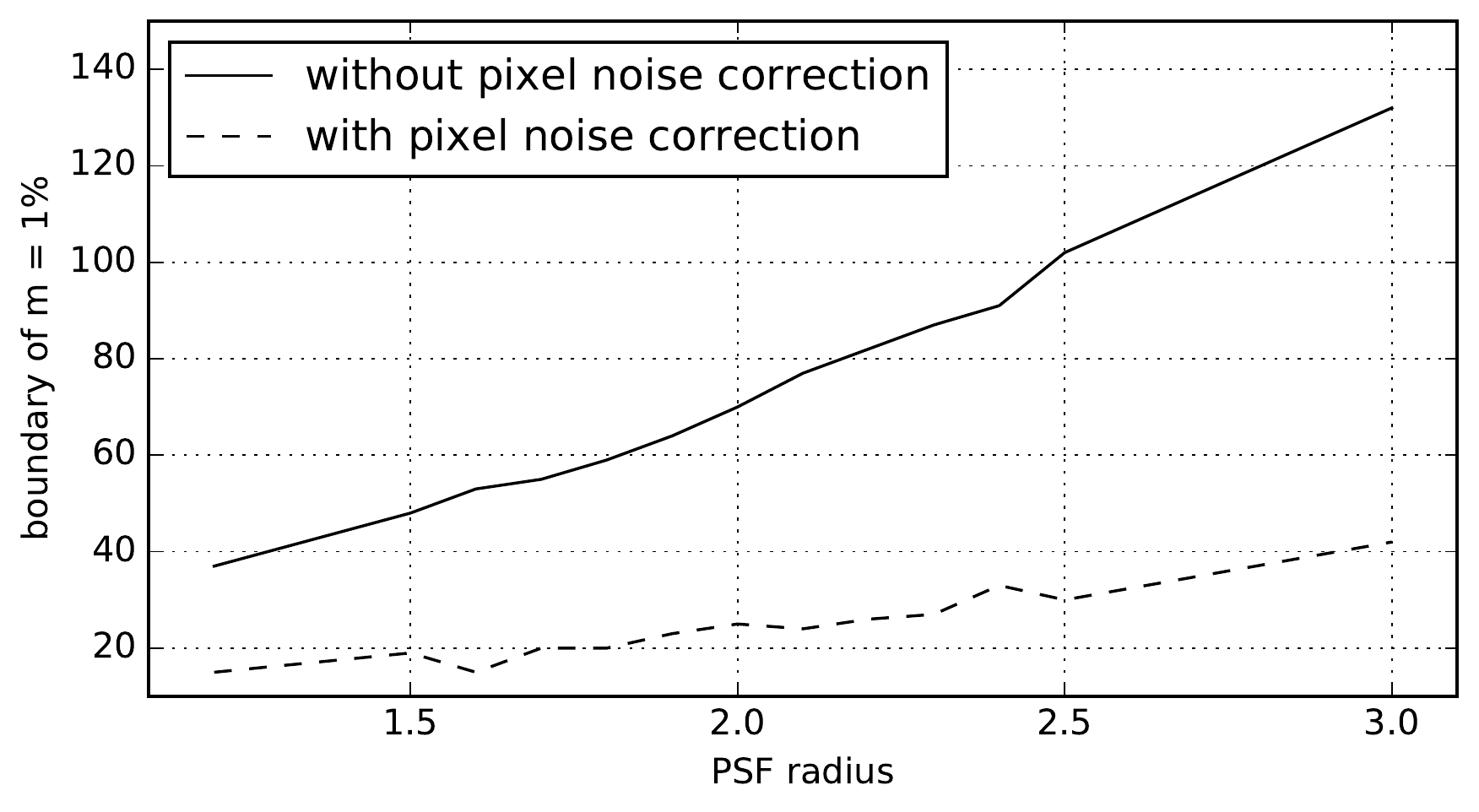}}
\caption{
\label{fig:SNRborderS}
same figure as figure \ref{fig:SNRborderG} except galaxy profile is Sersic.
}
\end{figure*}


\begin{figure*}[tbp]
\centering
\resizebox{1.0\hsize}{!}{\includegraphics{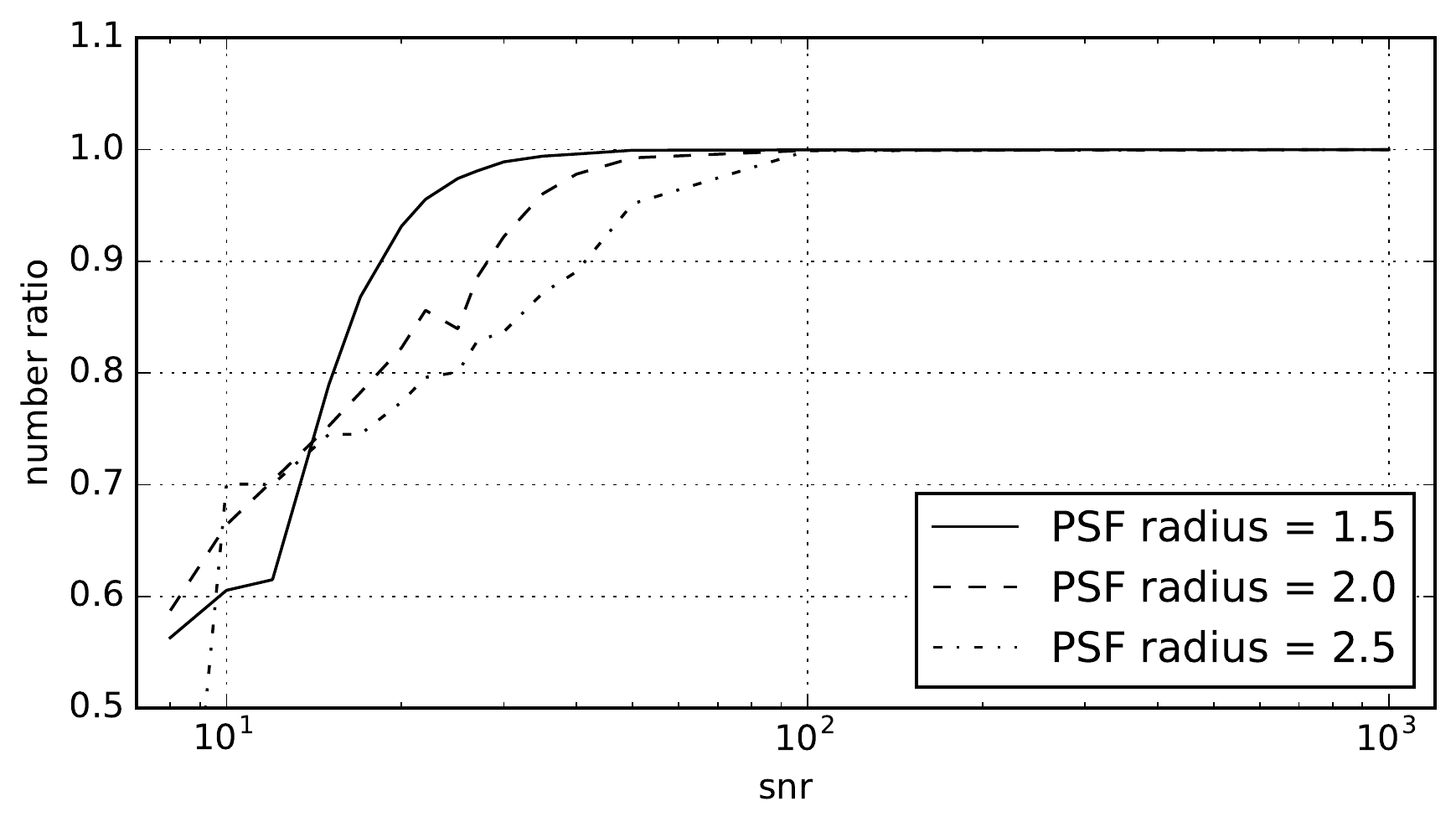}}
\caption{
\label{fig:reducednumberG}
The number ratio of the galaxy which can be used for measuring shear as a function of SNR.
1.0 means all of the galaxies are used for calculating the multiplicative bias.
PSF radius is 1.5(solid line), 2.0(dashed line) and 2.5(dot and dashed line) pixels.
}
\end{figure*}

\begin{figure*}[tbp]
\centering
\resizebox{1.0\hsize}{!}{\includegraphics{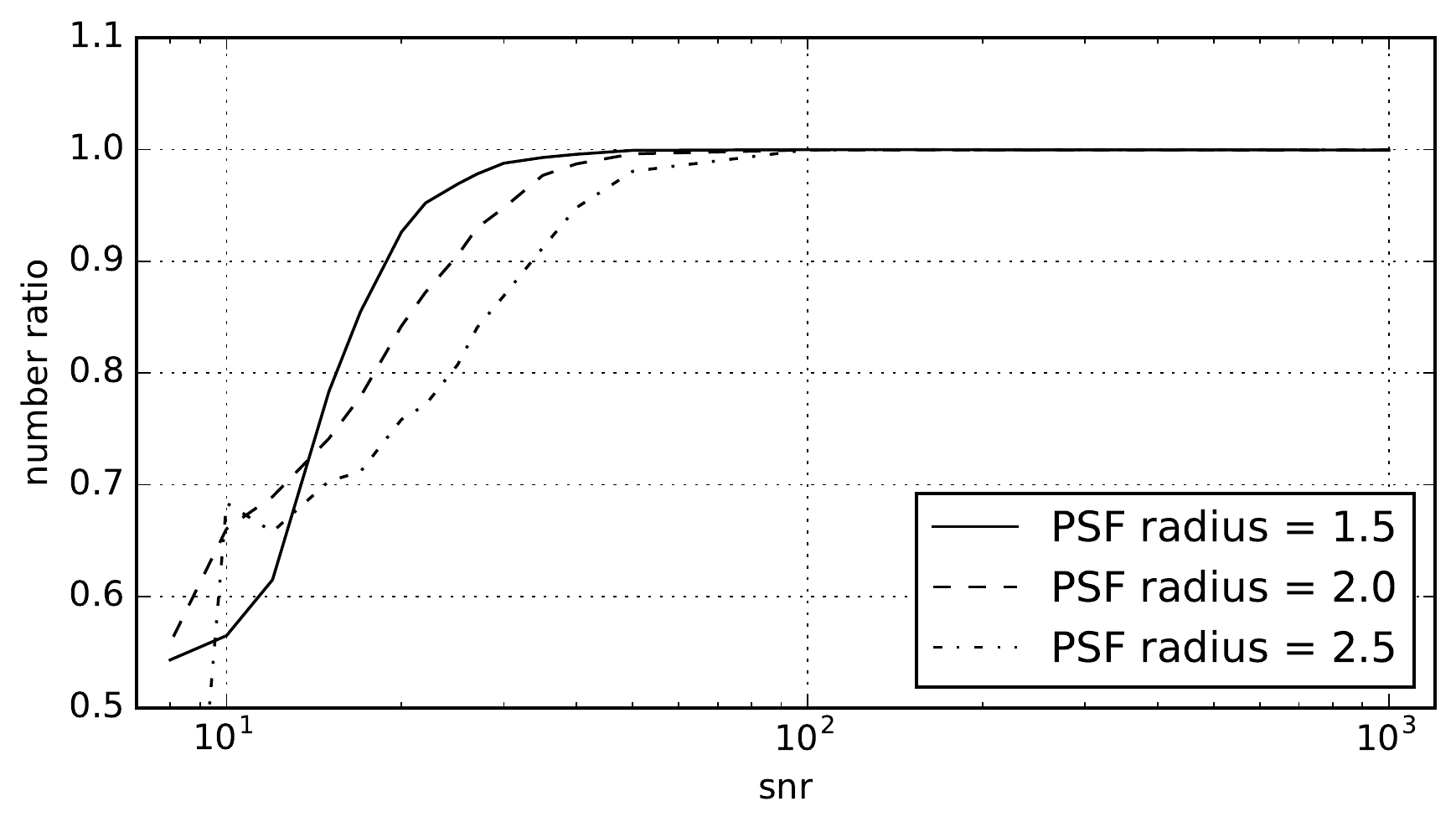}}
\caption{
\label{fig:reducednumberS}
Same figure as \ref{fig:reducednumberG} except galaxy profile is Sersic.
}
\end{figure*}

As the next test of our photon noise correction method, we use real galaxy image sets in GalSim(Rowe et al. 2015) taken by HST(Hubble Space Telescope).
We used 215336 number of galaxy images and add PSF which has the same size as the galaxy, then we added photon noise for the PSF smeared galaxy image.
The SN distribution of the galaxy is uniform between 5 and 100.
Next, we measured PSF corrected ellipticity and PSF+PN corrected ellipticity.
Then, we fit the measured ellipticity$\mbe$ by first order of true ellipticity $\mbe_{true}$ with 3$\sigma$ clipping as
\begin{eqnarray}
\mbe= \lr{m+1}\mbe_{true} + c,
\end{eqnarray}
where true ellipticity means the measured ellipticity without photon noise.
Figure 14 and 15 show the distribution of the difference between the measured ellipticity and the true 
ellipticity versus the true ellipticity without and with photon noise correction.
The results show 
\begin{eqnarray}
m_{meas} &=& \lr{0.9132-0.0978i}\times10^{-2}\\
c_{meas} &=& \lr{0.035-0.6389i}\times10^{-3}\\
m_{cor} &=& \lr{-0.1127+0.0233i}\times10^{-2}\\
c_{cor} &=& \lr{-0.1357-0.5597i}\times10^{-3},
\end{eqnarray}
where subscript ``meas''  and ``cor'' mean bias in PSF corrected ellipticity and PSF+PN corrected ellipticity.
We can see PSF corrected ellipticity has around 1\% noise bias, 
but PSF and photon noise corrected ellipticity has around 0.1\%,
so it is confirmed that the analytical correction improves the precision of weak lensing shear measurement.
\begin{figure*}[tbp]
\centering
\resizebox{1.0\hsize}{!}{\includegraphics{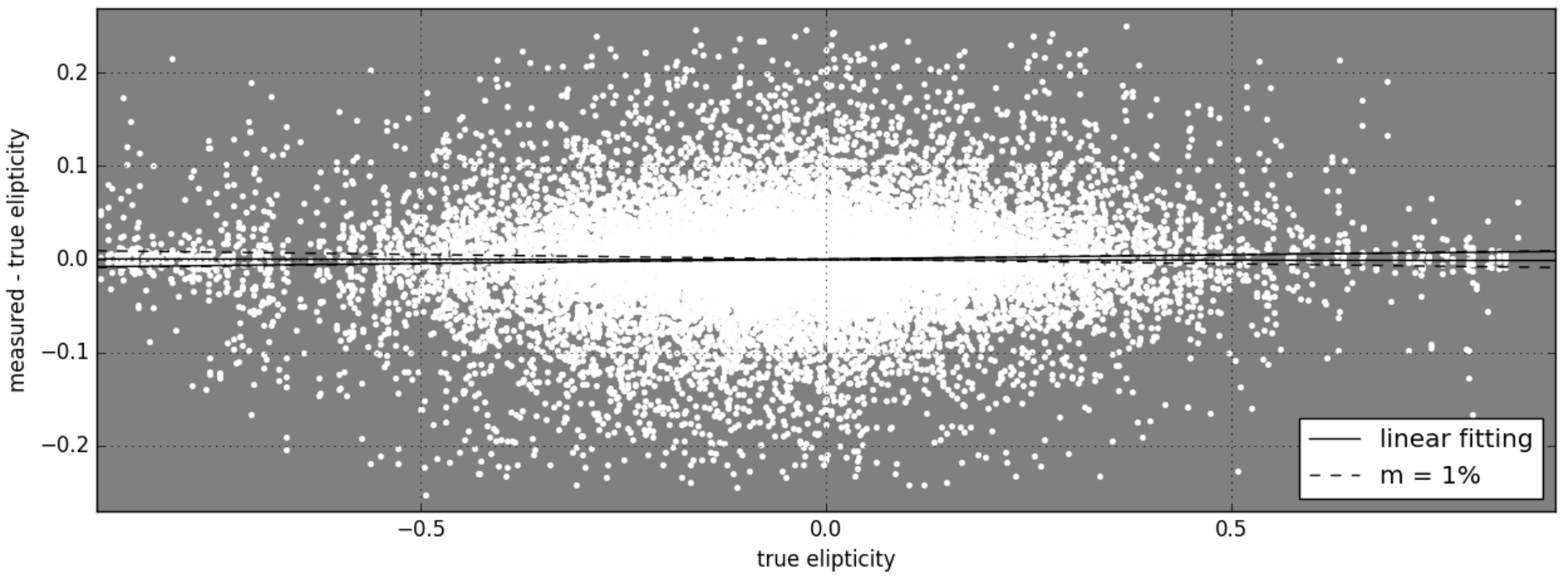}}
\caption{
\label{fig:GalSim_m}
 The distribution of the difference between the measured ellipticity and the true 
ellipticity versus the true ellipticity without photon noise correction. 
The white plots mean differences of measured ellipticity and true ellipticity, where both real and imaginal parts are plotted by the same white circles and we plot only about 10000 results of the 215336 results.
The black lines are linear fitting of the 215336 results(real and imaginary parts) and the black dashed line means 1\% multiplicative bias.
}
\end{figure*}
\begin{figure*}[tbp]
\centering
\resizebox{1.0\hsize}{!}{\includegraphics{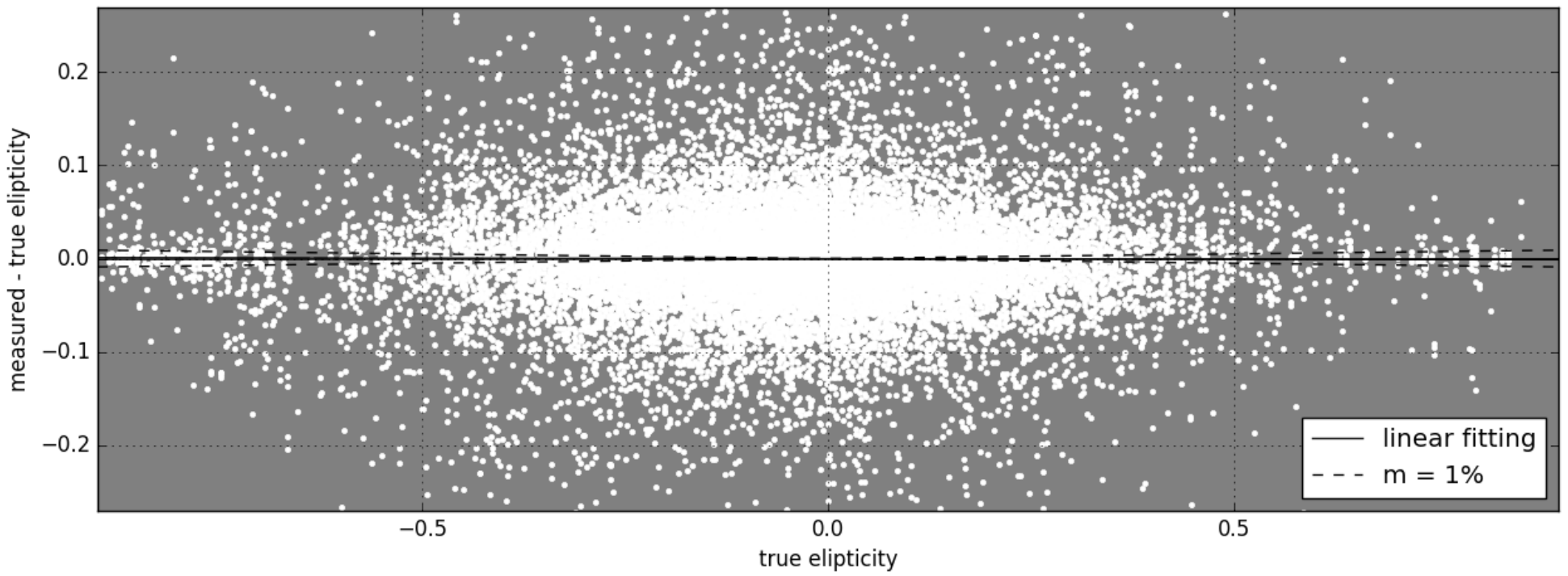}}
\caption{
\label{fig:GalSim_c}
Same figure as \ref{fig:GalSim_m} except the ellipticities are measured with PN correction.
}
\end{figure*}

\section{Summary}

In this paper, we have developed a new formulation of PSF correction within the framework of ``ERA'' to correct photon noise effect. 
The basic idea is very simple, adding photon noise in the equation of measuring PSF corrected ellipticity, then expanding the photon noise up to the 2nd order in the deviation from the true ellipticity with the assumption that the amplitude of the photon noise count is small.
The 1st order photon noise effect is random, so the mean value is 0 but it can be used to estimate the standard deviation of ellipticity from photon noise. 
The intrinsic ellipticity can be measured by the 1st photon noise effect with the assumption that the observed standard deviation has two components, and it is important to determine the weight of each galaxy.
The mean of the 2nd order photon noise gives the systematic error and is very important for precise weak lensing shear analysis.

From the result of the simulation test, 
it is confirmed that the noise bias obeys an inverse square law with respect to  SNR and the bias depends on the size ratio between galaxy and PSF in high SNR region.
In such region, our correction method can correct $80\%\sim90\%$ of the multiplicative bias.
In the low SNR region, however, the inverse square law is no more applicable, this would mean that the higher order effect is dominant in such region. The boundary at which the inverse square law begin to break down depend on the PSF size, namely,  if  the PSF size is larger, then SNR of the boundary is higher.   
As Fig. 12 and 13. shows, the noise bias after the correction by our method exceeds $1\%$ in such low SNR region.
How to correct the noise in the lower SNR region  is one of our feature works.
 
In the following paper, we will apply our method to HSC SSP wide field survey data 
 to estimate how much number density of galaxy in total and in each redshift bins can be increased by this correction method.

\section*{Acknowledgements}
We would thank Erin Shelden very much for many useful discussions. 
This work is supported in part by a Grant-in-Aid for Science Research from
JSPS(No.17K05453 to T.F).


\end{document}